\documentclass[lettersize,journal]{IEEEtran}
\IEEEoverridecommandlockouts
\usepackage{amsmath,amssymb,amsfonts}
\usepackage{algorithmic}
\usepackage{algorithm}
\usepackage{xcolor}
\usepackage{array}
\usepackage{subfigure}
\usepackage{color} 
\usepackage{setspace}
\usepackage{textcomp}
\usepackage{stfloats}
\usepackage{url}
\usepackage{verbatim}
\usepackage{graphicx}
\usepackage{cite}

\def\BibTeX{{\rm B\kern-.05em{\sc i\kern-.025em b}\kern-.08em
		T\kern-.1667em\lower.7ex\hbox{E}\kern-.125emX}}
\usepackage{balance}
\usepackage[colorlinks=true,linkcolor=red,citecolor=green]{hyperref}
\begin{document}
	\title{Two-Timescale Sum-Rate Maximization for Movable Antenna Enhanced Systems}
	\author{\IEEEauthorblockN{Xintai Chen, Biqian Feng, Yongpeng Wu, \emph{Senior Member, IEEE}, \\Derrick Wing Kwan Ng, \emph{Fellow, IEEE}, and Robert Schober, \emph{Fellow, IEEE}}
		\thanks{This paper has been presented in part at the 2023 IEEE Global Communications Conference (GLOBECOM), Kuala Lumpur, Malaysia \cite{JBaA}.}
		\thanks{X. Chen, B. Feng, and Y. Wu are with the Department of Electronic Engineering, Shanghai Jiao Tong University, Minhang 200240, China (e-mail: chenxintai@sjtu.edu.cn; fengbiqian@sjtu.edu.cn; yongpeng.wu@sjtu.edu.cn).}
		\thanks{D. W. K. Ng is with the School of Electrical Engineering and Telecommunications, University of New South Wales, Sydney, NSW, Australia (e-mail: w.k.ng@unsw.edu.au).}
		\thanks{R. Schober is with the Institute for Digital Communications, Friedrich-Alexander University Erlangen-N\"{u}rnberg (FAU), 91058 Erlangen, Germany (e-mail: robert.schober@fau.de).}
		\thanks{Corresponding author: Yongpeng Wu.}}
	
	\maketitle
	
	\begin{abstract}
		This paper studies a novel movable antenna (MA)-enhanced multiuser multiple-input multiple-output downlink system designed to improve wireless communication performance. We aim to maximize the average achievable sum rate through two-timescale optimization exploiting instantaneous channel state information at the receiver (I-CSIR) for receive antenna position vector (APV) design and statistical channel state information at the transmitter (S-CSIT) for transmit APV and covariance matrix design. We first decompose the resulting stochastic optimization problem into a series of short-term problems and one long-term problem. Then, a gradient ascent algorithm is proposed to obtain suboptimal receive APVs for the short-term problems for given I-CSIR samples. Based on the output of the gradient ascent algorithm, a series of convex objective/feasibility surrogates for the long-term problem are constructed and solved utilizing the constrained stochastic successive convex approximation (CSSCA) algorithm. Furthermore, we propose a planar movement mode for the receive MAs to facilitate efficient antenna movement and the development of a low-complexity primal-dual decomposition-based stochastic successive convex approximation (PDD-SSCA) algorithm, which finds Karush-Kuhn-Tucker (KKT) solutions almost surely. Our numerical results reveal that, for both the general and the planar movement modes, the proposed two-timescale MA-enhanced system design significantly improves the average achievable sum rate and the feasibility of the formulated problem compared to benchmark schemes.
	\end{abstract}
	
	\section{Introduction}
	Advanced multiple-input multiple-output (MIMO) techniques have been proposed to augment the efficacy of wireless communication systems by leveraging their available spatial degrees of freedom (DoFs) \cite{CLoM}. However, due to the fixed and uniformly spaced positions of antennas in conventional MIMO systems, it is impossible to fully exploit the spatial variations of the wireless channel across the entire transmit/receive region. To address this issue, two innovative techniques have been proposed: non-uniform antenna arrays \cite{Nlaa, Nadf2} and movable antennas (MAs) (a specific application of the concept of fluid antennas \cite{BLIF, FASN, FAwL}) \cite{Mapa, Mccf, JBaA, WSRM}. These advanced techniques can effectively harness otherwise untapped spatial diversity to strategically enhance the performance of MIMO systems by optimizing the physical placement of antennas.
	
	In the existing literature, non-uniform antenna arrays have been mainly exploited to mitigate the limitations in the available spatial DoFs arising from rank-deficient MIMO channels at millimeter wave or terahertz frequencies \cite{Nlaa, Nadf2}. However, the performance of non-uniform antenna arrays may not consistently surpass that of uniform arrays, since the positions of the antennas cannot dynamically adapt to rapidly varying wireless communication environments. As a remedy, fluid antennas have emerged as a promising alternative. Specifically, fluid antennas conceptually employ a radiating structure composed of software-controllable fluidic, conductive, or dielectric elements. These elements have the unique capacity to alter their shape, size, and/or position to reconfigure the polarization, operating frequency, radiation pattern, and other electromagnetic characteristics \cite{BLIF}.  The majority of the existing research on fluid antennas has focused on flexibly shifting the physical position of an antenna between a number of available ports, thereby establishing favorable channel conditions for efficient communication \cite{FASN, FAwL}.
	
	On the other hand, in MA-enhanced MIMO systems, a large number of antennas can be continuously repositioned, granting substantial flexibility for practical array design \cite{Mapa}. To unlock its potential, a few initial works studied the performance of MA-enhanced MIMO systems, mainly focusing on performance analysis \cite{Mapa}, antenna position vector (APV) optimization \cite{Mccf, JBaA, Eemf, WSRM, MAAE, MBFw, MEMC, MAEM, APaB, SRMf, MCwMZXLX, CMfF, MAEA, MAAEGQJK, SWCv}, and channel estimation \cite{CSBC, CEfM}. 
	For instance, in \cite{Mapa}, the maximum channel power gain achieved by a single receive MA was analyzed for both deterministic and stochastic channels. Besides, in \cite{Mccf}, an efficient alternating optimization method was developed to maximize the capacity of a point-to-point MIMO system. In particular, the proposed approach iteratively optimizes the transmit and receive APVs as well as the covariance matrix of the transmit signal. In addition, in \cite{JBaA}, the average achievable rate of a point-to-point MIMO system was maximized by jointly optimizing the APVs and the transmit covariance matrix exploiting statistical channel state information (CSI) through a constrained stochastic successive convex approximation (CSSCA) algorithm. In particular, the computational complexity for determining the solution was significantly reduced when restricting the movement of the MAs to separated regions. 
	Moreover, the application of enhanced multi-beamforming with a linear MA array has been investigated. For instance, in \cite{MAAE}, the optimal solutions for the APV and the antenna weight vector of a linear MA array were derived in closed form. These solutions were specifically tailored to manage a given number of MAs, achieving the maximum array gain for the desired direction while ensuring null steering across all undesired directions. Also, in \cite{MBFw}, a suboptimal iterative algorithm was proposed to maximize the minimum beamforming gain in multiple desired directions by optimizing both the APV and antenna weight vector. The design also ensured a limit on the maximum interference power in undesired directions.
	
	Multiuser communication systems employing MAs were investigated in \cite{MEMC, MAEM, APaB, SRMf, MCwMZXLX, CMfF}. In particular, in \cite{MEMC}, by exploiting perfect CSI, the total transmit power of multiple single-antenna users was minimized via two gradient descent methods employing zero-forcing and minimum mean-squared error criteria, respectively. Also, an iterative algorithm capitalizing on the generalized Bender’s decomposition was developed in \cite{MAEM} to determine the global-optimal solution for minimizing the total required transmit power to guarantee a certain minimum required signal-to-interference-plus-noise ratio for multiple downlink users by jointly optimizing the beamforming and APV at the base station (BS). 
	Moreover, in \cite{APaB}, the total downlink transmit power was minimized by jointly optimizing the beamforming matrices and APV taking into account a minimum signal-to-interference-plus-noise requirement for each user. This paper adopted an alternating optimization approach combined with the penalty method and successive convex approximation to address the nonconvex design problem. In addition, in \cite{SRMf}, the sum-rate of a multiuser downlink multiple-input single-output system was enhanced via the joint optimization of the transmit beamforming and APVs through a computationally efficient algorithm capitalizing on fractional programming, alternating optimization, and gradient descent methods. Furthermore, in \cite{MCwMZXLX}, a two-loop iterative algorithm combining particle swarm optimization and block coordinate descent method was proposed to maximize the minimum rate of multiple users in an uplink system, where the APVs, the receive combining at the BS, and the transmit power of the users were jointly optimized. On the other hand, by jointly optimizing the transmit covariance matrices and the APVs of the users, the maximization of the multiple access system capacity was studied in \cite{CMfF}, where closed-form bounds and approximations for the maximum capacity were derived. 
	
	Despite various research efforts, the performance of MIMO systems is still severely affected by the accuracy of the instantaneous CSI at the transmitter (I-CSIT) \cite{CSTC}. However, it is difficult to acquire perfect I-CSIT, especially in frequency-division duplex systems, due to quantization errors. In particular, in frequency-division duplex systems, the receiver needs to first estimate the CSI and then feed back a quantized version to the transmitter. Besides, to achieve full multiplexing gain, the feedback rate must scale linearly with $\log_2\text{SNR}$ \cite{CSTC,MBCW}. As a result, systems operating even at moderate SNR levels will require an unaffordable feedback overhead. Fortunately, obtaining statistical CSI at the transmitter (S-CSIT), e.g., the spatial correlation and channel mean, is generally feasible through long-term feedback or covariance extrapolation \cite{FmMv}, since it tends to remain invariant over a much longer period of time. Therefore, when I-CSIT is not available, exploiting S-CSIT for resource allocation design serves as a practical alternative for implementing MA systems. On the other hand, since instantaneous CSI can be acquired at the receiver directly, basing the receive design on instantaneous CSI at the receiver (I-CSIR) is realistic for low-mobility user terminals (UTs). Motivated by the above discussion, assuming availability of S-CSIT and I-CSIR, this paper investigates the resource allocation design in multiuser MIMO downlink systems equipped with MAs at the BS and UTs. The resulting optimization problem is very challenge since it is a nonconvex stochastic optimization problem involving two-timescale variables. In fact, the performance under this realistic setting achievable by the joint optimization of the transmit design based on S-CSIT and the receive APVs based on I-CSIR has not been explored in existing works, e.g., \cite{Mapa, Mccf, JBaA, MAAE, MBFw, MEMC, MAEM, APaB, SRMf, MCwMZXLX, CMfF, MAEA, MAAEGQJK, SWCv}. We note furthermore that the design of receive APVs exploiting I-CSIR was not considered in the conference version of this paper \cite{JBaA}. The main contributions of this paper can be summarized as follows:
	\begin{itemize}
		\item [1)]
		Firstly, we investigate a two-timescale design of MA-enhanced multiuser MIMO systems exploiting both S-CSIT and I-CSIR. To this end, we adopt the field-response based channel model to characterize the relationship between the channel matrices and the APVs. 
		\item [2)]
		Secondly, we develop a two-timescale joint beamforming and APV optimization algorithm framework to maximize the average achievable sum rate for the case where the transmit and receive MAs can move freely within given areas, which we refer to as the general movement mode (GMM). The non-convex constraints associated with the receive APVs are an obstacle to efficiently obtaining a Karush-Kuhn-Tucker (KKT) solution by, e.g., a primal-dual decomposition-based stochastic successive convex approximation (PDD-SSCA) algorithm. Therefore, we focus on acquiring a suboptimal solution by directly decomposing the resulting two-timescale stochastic optimization problem into a series of short-term problems and one long-term problem. Then, a gradient ascent (GA) algorithm is proposed to obtain suboptimal solutions for the short-term problems for given I-CSIR samples. Based on the output of the GA algorithm, a series of convex objective/feasibility surrogates for the long-term problem are constructed and solved utilizing the CSSCA algorithm.
		\item [3)]
		Thirdly, we constrain the movement of the receive MAs, which we refer to as the planar movement mode (PMM), to increase the efficiency of the antenna movement. Moreover, the PMM facilitates the development of an efficient PDD-SSCA algorithm, which provides a KKT solution of the resulting two-timescale stochastic optimization problem almost surely. In particular, in each iteration, the proposed PDD-SSCA algorithm first executes a gradient projection (GP) algorithm to acquire stationary points of the short-term problems associated with a mini-batch of I-CSIR. Then, based on the output of the short-term algorithm, a convex objective/feasibility surrogate for the long-term problem is constructed and solved exploiting the CSSCA algorithm. 
		\item [4)]
		Finally, we provide extensive numerical results to demonstrate the significant gains in the average achievable sum rate and the feasibility of the formulated problem realized with the proposed MA-enhanced MIMO system compared to several benchmarks, including MIMO systems employing conventional fixed-position uniform planar arrays (UPAs). For all considered simulation settings, the proposed GMM scheme delivers the best achievable rate performance.
	\end{itemize}
	
	The remainder of this paper is organized as follows. Section II presents the system model and problem formulation. Section III provides the proposed two-timescale optimization framework for the GMM. Section IV introduces the PMM for the receive MAs and the PDD-SSCA algorithm. Numerical results and corresponding discussions are presented in Section V. Finally, Section VI concludes this paper.
	
	\textbf{Notations}: Vectors (lower case) and matrices (upper case) are presented in boldface. $(\cdot)^{T}$ and $(\cdot)^{H}$ denote the transpose and conjugate transpose (Hermitian), respectively. $[\mathbf{A}]_{p_1,\cdots,p_K}$ denotes the entry with index $p_1,\cdots,p_K$ of $K$-dimensional tensor $\mathbf{A}$. The ensemble expectation, matrix trace, and determinant operations are
	denoted by $\mathbb{E}\{\cdot\}$, $\operatorname{tr}(\cdot)$, and $\operatorname{det}(\cdot)$, respectively. $\mathbf{A} \succeq \mathbf{0}$ indicates that $\mathbf{A}$ is a positive semi-definite matrix. $\operatorname{diag}\{\mathbf{a}\}$ returns a diagonal matrix with the $i$-th main diagonal entry equal to the $i$-th entry of vector $\mathbf{a}$. $\operatorname{diag}\{\mathbf{A}\}$ returns a vector with the $i$-th entry equal to the $i$-th main diagonal entry of matrix $\mathbf{A}$. $\operatorname{Re}\{\mathbf{A}\}$ returns a real-valued matrix whose entries equal to the real parts of the entries of matrix $\mathbf{A}$.
	We adopt $\mathbf{I}_K$ to represent the $K$-by-$K$ identity matrix. $\left\|\mathbf{a}\right\|$ and $\left\|\mathbf{A}\right\|$ denote the $L_2$-norm of vector $\mathbf{a}$ and the Frobenius norm of matrix $\mathbf{A}$, respectively. $\left[\mathbf{x}\right]_{\mathcal{X}}$ projects $\mathbf{x}$ into domain $\mathcal{X}$. 
	$\mathcal{CN}\left(\mathbf{0}, \mathbf{\Gamma}\right)$ denotes the circularly symmetric complex Gaussian distribution with mean $\mathbf{0}$ and covariance matrix $\mathbf{\Gamma}$.
	
	\section{System Model and Problem Formulation}
	\label{subsection: MA-Enhanced MIMO System}
	\begin{figure*}[t]
		\centering
		\includegraphics[width=0.95\textwidth,height=0.37\textwidth]{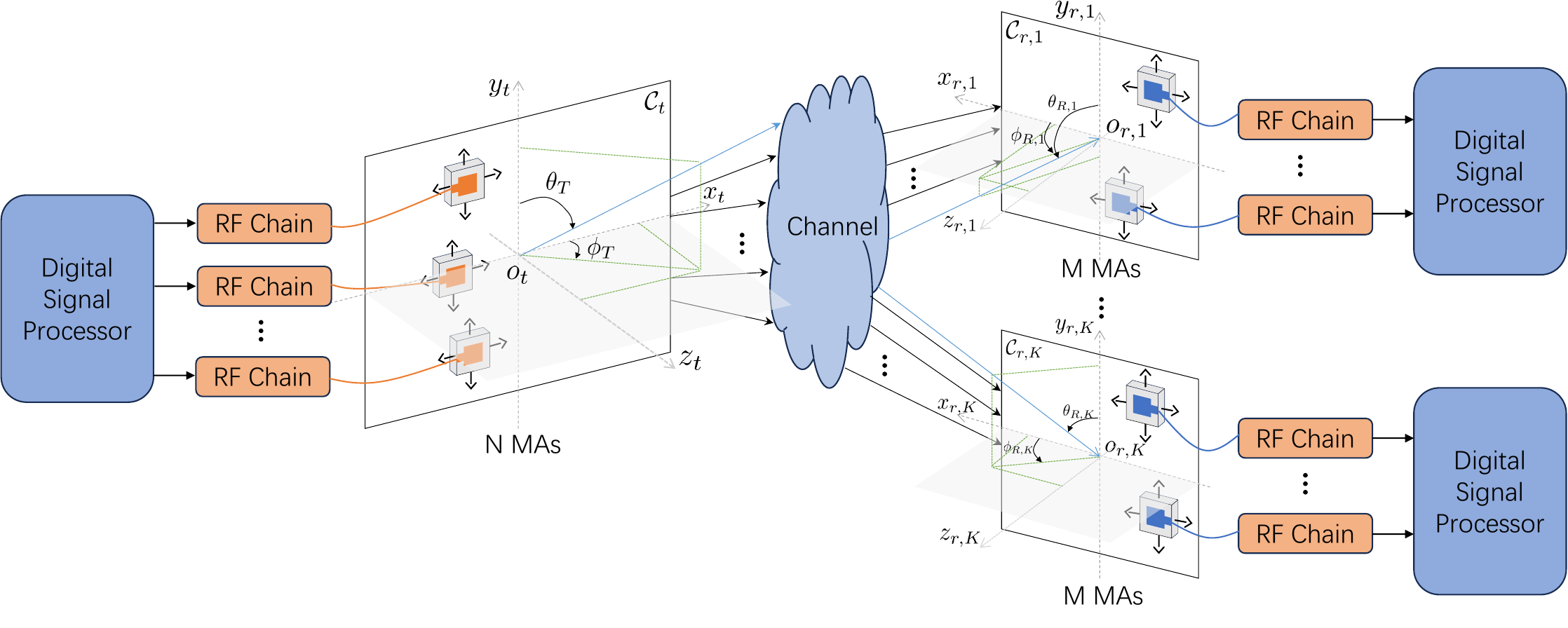}
		\caption{Illustration of the downlink transmission from the BS to $K$ UTs. All transceivers are equipped with MAs.}
		\label{fig1}
	\end{figure*}
	As shown in Fig.~\ref{fig1}, we consider a typical multiuser MA-enhanced MIMO downlink communication system, where the BS is equipped with $N$ MAs to simultaneously serve $K$ low-mobility UTs, each of which is equipped with $M$ MAs. 
	Assume that all transmit MAs at the BS and all receive MAs at UT $k \in\{ 1,\cdots,K\}$ can move within certain common regions $\mathcal{C}_t$ and $\mathcal{C}_{r,k}$, respectively, i.e., the position of transmit antenna $n$ at the BS satisfies $\mathbf{t}_n \triangleq \left(x_{t,n},y_{t,n}\right) \in \mathcal{C}_t$ and the position of receive antenna $m$ at UT $k$ satisfies $\mathbf{r}_{k,m} \triangleq \left(x_{r,k,m},y_{r,k,m}\right) \in \mathcal{C}_{r,k}$. By stacking the coordinates of all antennas together, the transmit APV at the BS and the receive APV at UT $k$ are denoted by $\mathbf{t}=\left(\mathbf{t}_1, \mathbf{t}_2, \cdots, \mathbf{t}_N\right)^{T}$ and $\mathbf{r}_k=\left(\mathbf{r}_{k,1}, \mathbf{r}_{k,2}, \cdots, \mathbf{r}_{k,M}\right)^{T}$, respectively. To avoid potential coupling between adjacent MAs, a minimum distance $D \geq \lambda/2$ is required between each pair of MAs \cite{AAfL}, i.e., $\left\|\mathbf{t}_i-\mathbf{t}_j\right\| \geq D, \forall i \neq j$, and $\left\|\mathbf{r}_{k,i}-\mathbf{r}_{k,j}\right\| \geq D, \forall k$, $\forall i \neq j$, where $\lambda$ is the wavelength of the signal carrier. Then, the received signal at UT $k$ is given by
	\begin{equation}
		\mathbf{y}_k=\mathbf{H}_k(\mathbf{t},\mathbf{r}_k) \sum_{i=1}^K \mathbf{x}_i+\mathbf{z}_k, \forall k,
	\end{equation}
	where $\mathbf{H}_k\left(\mathbf{t},\mathbf{r}_k\right) \in \mathbb{C}^{M \times N}$ represents the channel matrix between the BS and UT $k$, which depends on APVs $\mathbf{t}$ and $\mathbf{r}_k$. The signal intended for UT $k$, $\mathbf{x}_k$, follows a zero-mean circularly symmetric complex Gaussian distribution with covariance matrix $\mathbf{Q}_k \triangleq \mathbb{E}\left\{\mathbf{x}_k \mathbf{x}_k^{H}\right\}$ and satisfies $\mathbb{E}\left\{\mathbf{x}_k \mathbf{x}_{k'}^{H}\right\}=\mathbf{0}$, $\forall k' \neq k$. $\mathbf{z}_k \sim \mathcal{CN}\left(\mathbf{0}, \sigma^2 \mathbf{I}_{M}\right)$ denotes the circularly symmetric complex Gaussian noise with mean $\mathbf{0}$ and variance $\sigma^2\mathbf{I}_{M}$ at UT $k$. For given channel $\mathbf H_k(\mathbf t,\mathbf r_k)$, the achievable rate for UT $k$ is given by
	\begin{equation}
		\begin{aligned}
			&R_k\left(\mathbf{t},\mathbf{r}_k,\left\{\mathbf{Q}_i\right\},\mathbf{H}_k\right) \triangleq \\& \log _2 \operatorname{det} \left( \mathbf{I}_{M}+\mathbf{H}_k \mathbf{Q}_k \mathbf{H}_k^{H} \left( \sigma^2 \mathbf{I}_{M} + \mathbf{H}_k \sum_{i\neq k}\mathbf{Q}_{i} \mathbf{H}_k^{H} \right)^{-1} \right).
		\end{aligned}
	\end{equation}
	
	\subsection{Field-Response Based Channel Model}
	\label{subsection: Channel Model}
	We assume that all UTs are in the far field of the BS such that the angles of departure (AoDs) and angles of arrival (AoAs) for different positions in transmit region $\mathcal{C}_t$ and receive region $\mathcal{C}_{r,k}$ are identical, respectively \cite{Mapa,Mccf,JBaA}. We assume that there are $L_t$ transmit paths and $L_r$ receive paths in the channel of each UT. Let $(\theta_{t,k}^{l_t}, \phi_{t,k}^{l_t})$ and $(\theta_{r,k}^{l_r}, \phi_{r,k}^{l_r})$ denote the elevation and azimuth angles of the $l_t$-th transmit path and the $l_r$-th receive path between the BS and UT $k$, respectively. Then, the differences in the signal propagation distance between position $\mathbf{t}_n$ and the origin of transmit region $\mathcal{C}_t$ at the BS and that between position $\mathbf{r}_{k,m}$ and the origin of receive region $\mathcal{C}_{r,k}$ at UT $k$ are, respectively, given by
	\begin{equation}
		\begin{aligned}
			\rho_{t,k}^{l_t}(\mathbf{t}_n)&=x_{t,n} \sin \theta_{t,k}^{l_t} \cos \phi_{t,k}^{l_t}+y_{t,n} \cos \theta_{t,k}^{l_t},\\
			\rho_{r,k}^{l_r}(\mathbf{r}_{k,m})&=x_{r,k,m} \sin \theta_{r,k}^{l_r} \cos \phi_{r,k}^{l_r}+y_{r,k,m} \cos \theta_{r,k}^{l_r}.
		\end{aligned}
	\end{equation}
	Then, the transmit and receive field response vectors of MA $n$ at the BS and MA $m$ at UT $k$ can be, respectively, written as follows
	\begin{equation}
		\begin{aligned}
			\mathbf{g}_k(\mathbf{t}_n) \triangleq & \left[e^{j \frac{2 \pi}{\lambda} \rho_{t,k}^{1}(\mathbf{t}_n)}, \cdots, e^{j \frac{2 \pi}{\lambda} \rho_{t,k}^{L_t}(\mathbf{t}_n)}\right]^{T},\\
			\mathbf{f}_{k}(\mathbf{r}_{k,m}) \triangleq & \left[e^{j \frac{2 \pi}{\lambda} \rho_{r,k}^{1}(\mathbf{r}_{k,m})},\cdots, e^{j \frac{2 \pi}{\lambda} \rho_{r,k}^{L_r}(\mathbf{r}_{k,m})}\right]^{T}.
		\end{aligned} 
	\end{equation}
	Furthermore, the channel matrix $\mathbf{H}_k(\mathbf{t},\mathbf{r}_k)$ between the BS and UT $k$ can be written as follows
	\begin{equation}
		\label{eq: discrete channel matrix}
		\begin{aligned}	
			\mathbf{H}_k(\mathbf{t},\mathbf{r}_k)
			= \mathbf{F}_{k}^H\left(\mathbf{r}_k\right) \mathbf{\Sigma}_k \mathbf{G}_{k}\left(\mathbf{t}\right),
		\end{aligned}
	\end{equation}
	where $\mathbf{F}_k(\mathbf{r}_k) \triangleq\left[\mathbf{f}_k\left(\mathbf{r}_{k,1}\right), \cdots, \mathbf{f}_k\left(\mathbf{r}_{k,M}\right)\right]$ and $\mathbf{G}_k(\mathbf{t}) \triangleq\left[\mathbf{g}_k\left(\mathbf{t}_1\right), \cdots, \mathbf{g}_k\left(\mathbf{t}_N\right)\right]$ denote the field response matrices at the BS and UT $k$, respectively, and $\mathbf{\Sigma}_k$ is the path-response matrix. 
	
	In MA-enhanced systems, acquiring I-CSIT at the BS is usually prohibitively expensive due to the continuous dynamic variations in the physical environment, especially in frequency-division duplex systems. Thus, in this paper, the slowly-changing S-CSIT is exploited for the transmit design of the considered MA-enhanced system. To this end, we assume that the path gains in $\mathbf{\Sigma}_k$ are independently and identically distributed (i.i.d.), and modeled as zero-mean Gaussian random variables \cite{FAAM}. 
	
	\subsection{Two-Timescale Transmission Framework}
	In this paper, we aim to maximize the average achievable sum rate with respect to (w.r.t.) covariance matrices $\{\mathbf Q_i\}$, transmit APV $\mathbf t$, and receive APVs $\{\mathbf r_k\}$ based on a two-timescale transmission framework. The average achievable sum rate is considered since we assume that the BS has only access to statistical CSI via CSI feedback from the UTs. In contrast, UT $k$ can acquire I-CSIR knowledge of its channel to the BS via downlink pilot signals, along with  the aggregate instantaneous interference-plus-noise covariance matrix, based on a compressed sensing-based method \cite{CSBC, CEfM}. Specifically, the BS transmits pilots to the UTs and then each UT estimates its I-CSIR in each coherence time interval. After a few coherence time intervals, the statistical CSI of the respective channel can be accurately estimated by each UT based on its previous I-CSIR estimates. This piece of information is then fed back to the BS as S-CSIT for transmit APV and covariance matrix design. Subsequently, the BS transmits the solution obtained for the transmit APV and covariance matrices to the UTs. Assuming low-mobility UTs, the coherence time is sufficiently long to enable the optimization of the receive APV for each UT based on its own I-CSIR.
	
	We take into account the constraints on the average achievable rate of each UT, the antenna positions, and the available transmit power. Accordingly, the proposed optimization problem is formulated as follows
	\begin{equation}
		\label{op: original 1}
		\begin{aligned}
			\max _{\mathbf{t}\in \mathcal{C}_t, \left\{\mathbf{Q}_i\succeq\mathbf 0\right\}} \quad &\sum_{k=1}^K \bar R_k\left(\mathbf{t},\left\{\mathbf{Q}_i\right\}\right) &\\
			\text { s.t. } \quad\quad
			&\text{C1$_k$: }\bar R_k\left(\mathbf{t},\left\{\mathbf{Q}_i\right\}\right) \geq R_{\mathrm{min}}, &\forall k, \\
			&\text{C2: }\left\|\mathbf{t}_i-\mathbf{t}_j\right\|^2 \geq D^2,  \forall i \neq j,\\
			&\text{C3$_k$: }\left\|\mathbf{r}_{k,i}-\mathbf{r}_{k,j}\right\|^2 \geq D^2, \forall i \neq j, &\forall k, \\
			&\text{C4: }\sum_{i=1}^K\operatorname{tr}(\mathbf{Q}_i) \leq P,
		\end{aligned}
	\end{equation}
	where the achievable rate $\bar R_k\left(\mathbf{t},\left\{\mathbf{Q}_i\right\}\right)$ is defined as
	\begin{equation}
		\begin{aligned}
			\label{eq: Rbar}
			\bar R_k\left(\mathbf{t},\left\{\mathbf{Q}_i\right\}\right) \triangleq \mathbb{E}_{\mathbf{H}_k} \left\{ \max_{\left\{\tilde{\mathbf{r}}_k\in \mathcal{C}_{r,k}\right\}} R_k\left(\mathbf{t},\tilde{\mathbf{r}}_k,\left\{\mathbf{Q}_i\right\},\mathbf{H}_k\right) \right\}.
		\end{aligned}
	\end{equation}
	$P \geq 0$ is the given maximum available transmit power and $R_{\mathrm{min}}$ is the required minimum rate of each UT. Here, $\tilde{\mathbf{r}}_k=\left(\mathbf{r}_{k,1}, \mathbf{r}_{k,2}, \cdots, \mathbf{r}_{k,M}\right)^{T}$ represents the short-term receive APV optimized for I-CSIR $\mathbf{H}_k$ in each coherence time interval.
	
	Note that obtaining even a KKT solution to problem \eqref{op: original 1} is challenging for the following reasons. Firstly, the achievable rates in the objective function and constraints C1$_k$ are not only highly nonconvex in terms of the APVs and transmit covariance matrices, but involve long-term variables $\mathbf{t}$ and $\left\{\mathbf{Q}_i\right\}$ as well as short-term variables $\{\tilde{\mathbf{r}}_k\}$. It is generally challenging if not impossible to derive the exact expected value of the average achievable rates in \eqref{op: original 1} in closed form. Moreover, the position constraints of the receive MAs, C3$_k$, $\forall k$, are nonconvex. This is an obstacle for developing a computationally efficient solution for extraction of a stationary point for the short-term problems and the gradient information that is necessary for the PDD-SSCA algorithm \cite{TSSO}. As a compromise, we focus on acquiring a suboptimal solution by directly incorporating the GA method into the CSSCA algorithm \cite{SSCA} in the following section. Besides, in Section \ref{section: Antenna Move Mode Design}, we propose a PDD-SSCA algorithm to obtain a KKT solution for the case, where the movements of the receive MAs are restricted to non-overlapping regions.
	
	\section{Design of MA-Enhanced MIMO System}
	\label{section: Design of MA-Enhanced MIMO System}
	Based on the above discussions, we first decompose problem \eqref{op: original 1} into a series of short-term problems and one stochastic long-term problem. 
	Specifically, the $K$ short-term problems for a given I-CSIR sample are given by
	\begin{equation}
		\label{op: decouple S} 	
		\begin{aligned}
			\max _{\tilde{\mathbf{r}}_k\in \mathcal{C}_{r,k}} \quad 	&R_k\left(\mathbf{t},\tilde{\mathbf{r}}_k,\left\{\mathbf{Q}_i\right\},\mathbf{H}_k\right) \\
			\text{s.t.} \quad
			&\text{C3}_k.
		\end{aligned}
	\end{equation}
	To mitigate the high computational complexity associated with acquiring stationary point solutions, a low-complexity GA algorithm is developed to obtain suboptimal solutions of these short-term problems. On the other hand, given long-term variables $\mathbf{t}$ and $\left\{\mathbf{Q}_i\right\}$, the receive APV for the long-term problem is set to the suboptimal solution $\tilde{\mathbf{r}}_k^{\hat{\ell}}\left(\mathbf{t},\left\{\mathbf{Q}_i\right\},\mathbf{H}_k\right)$ obtained for one I-CSIR sample by executing the GA algorithm for a sufficiently large number of iterations $\hat{\ell}$.
	Accordingly, the long-term problem is written as
	\begin{equation}
		\label{op: decouple L}
		\begin{aligned}
			\max _{\mathbf{t}\in \mathcal{C}_t, \left\{\mathbf{Q}_i\succeq\mathbf 0\right\}} 	\,\, &\sum_{k=1}^K \mathbb{E}_{\mathbf{H}_k} \left\{ R_k\left(\mathbf{t},\tilde{\mathbf{r}}_k^{\hat{\ell}},\left\{\mathbf{Q}_i\right\},\mathbf{H}_k\right)  \right\} \\
			\text { s.t. } \quad
			&\text{C1$'_k$},\forall k,\text{C2},\text{C4},
		\end{aligned}
	\end{equation}
	where constraint $\mathbb{E}_{\mathbf{H}_k} 	\left\{ R_k\left(\mathbf{t},\tilde{\mathbf{r}}_k^{\hat{\ell}},\left\{\mathbf{Q}_i\right\},\mathbf{H}_k\right)  \right\} \geq R_{\mathrm{min}}$ is denoted as C1$'_k$.
	From the literature \cite{SSCA}, it is known that the CSSCA method can handle stochastic optimization problems involving expectations over random system states even if the objective function and the constraints are nonconvex. Therefore, we resort to the CSSCA methodology for solving the stochastic problem in \eqref{op: decouple L} efficiently.
		
	\subsection{Short-term Receive APV Design Based on I-CSIR}
	Note that various general iterative algorithm frameworks, such as successive convex approximation and majorization-minimization algorithms, can be adopted to obtain a stationary point solution for nonconvex problems \cite{CO}, \cite{MMAi}. However, due to the highly nonconvex nature of the objective function in \eqref{op: decouple S} w.r.t. $\tilde{\mathbf{r}}_k$, it is challenging to find an approximate surrogate function that can closely approximate the objective function and thus achieve fast convergence. Moreover, even if an accurate approximation is found, the resulting subproblem may be still of high computational complexity \cite{Mccf}. Therefore, to address this nonconvex problem, we employ the GA method to update $\tilde{\mathbf{r}}_k$, which reduces the computational complexity. 
	
	We iteratively optimize the position of each receive MA while fixing all other MAs. In particular, the position of the $m$-th MA in iteration $\hat{\ell}$ is updated by moving along the gradient direction \cite{Aito}, i.e.,
	\begin{equation}
		\label{eq: RMA update}
		\mathbf{r}_{k,m}^{\hat{\ell}} = \left[ \mathbf{r}_{k,m}^{\hat{\ell}-1}+ \tau^{\hat{\ell},m} \nabla_{\mathbf{r}_{k,m}} R_k\left(\mathbf{t}, \tilde{\mathbf{r}}_{k}^{\hat{\ell}-1,m},\left\{\mathbf{Q}_i\right\},\mathbf{H}_k\right) \right]_{\mathcal{C}_{r,k}},
	\end{equation}
	where $\tilde{\mathbf{r}}_{k}^{\hat{\ell}-1,m}\triangleq \left(\mathbf{r}_{k,1}^{\hat{\ell}},\cdots,\mathbf{r}_{k,m-1}^{\hat{\ell}},\mathbf{r}_{k,m}^{\hat{\ell}-1},\cdots,\mathbf{r}_{k,M}^{\hat{\ell}-1}\right)^T$, since the $i$-th MA ($i< m$) has undergone $\hat{\ell}$ iterations, while the $j$-th MA ($j\geq m$) has undergone only $\hat{\ell}-1$ iterations, and $\tau^{\hat{\ell},m}$ is the step size selected by the backtracking line search for gradient ascent in iteration $\hat{\ell}$ \cite{CO}. In each iteration, we start with a large positive step size, $\tau^{\hat{\ell},m}=s$, and repeatedly reduce it to $\tau\tau^{\hat{\ell},m}$ with a factor $\tau\in(0,1)$, until the Armijo–Goldstein condition in \eqref{eq: Armijo–Goldstein condition} and the antenna distance constraints, i.e., $\left\|\mathbf{r}_{k,i}-\mathbf{r}_{k,m}\right\|^2 \geq D^2$, $\forall i\neq m$ are satisfied:
	\begin{equation}
		\begin{aligned}
			R_k&\left(\mathbf{t},\tilde{\mathbf{r}}_{k}^{\hat{\ell}',m'},\left\{\mathbf{Q}_i\right\},\mathbf{H}_k \right) \geq R_k\left(\mathbf{t},\tilde{\mathbf{r}}_{k}^{\hat{\ell}-1,m},\left\{\mathbf{Q}_i\right\},\mathbf{H}_k \right) \\&
			+ \xi \tau^{\hat{\ell},m} \left\|\nabla_{\mathbf{r}_{k,m}} R_k\left(\mathbf{t}, \tilde{\mathbf{r}}_{k}^{\hat{\ell}-1,m},\left\{\mathbf{Q}_i\right\},\mathbf{H}_k\right)\right\|^2, \label{eq: Armijo–Goldstein condition}
		\end{aligned}
	\end{equation}
	where $m'=m+1$, $\hat{\ell}'=\hat{\ell}-1$ when $m=1,2,\cdots,M-1$ and $m'=1$, $\hat{\ell}'=\hat{\ell}$ when $m=M$. $\xi \in (0, 1)$ is a given control parameter to guarantee that  the objective function achieves an adequate increase with the current step size. The overall GA algorithm terminates when the increment of the objective value over two consecutive iterations is less than a small positive value $\epsilon$. Note that if all $M$ MAs are updated simultaneously rather than iteratively in each iteration, a few active antenna distance constraints, $\left\|\mathbf{r}_{k,i}-\mathbf{r}_{k,m}\right\|^2 \geq D^2$, may cause zero increment in the objective and premature termination of the algorithm when the gradient points outwards the feasible region.
	
	The gradients $\nabla_{\mathbf{r}_{k,m}} R_k\left(\mathbf{t}, \tilde{\mathbf{r}}_{k}^{\hat{\ell}-1,m},\left\{\mathbf{Q}_i\right\},\mathbf{H}_k\right)$, $\forall m$, $\forall k$, required in \eqref{eq: RMA update} and \eqref{eq: Armijo–Goldstein condition} can be calculated according to \eqref{eq: rate differential r} in Appendix \ref{Ap: gradient rate}. In calculating these gradients, the major computational complexity is contributed by the matrix inversions $\left( \sigma^2 \mathbf{I}_{M}+\mathbf{H}_k \sum_{i=1}^K\mathbf{Q}_i \mathbf{H}_k^{H} \right)^{-1}$ and $\left( \sigma^2 \mathbf{I}_{M}+\mathbf{H}_k \sum_{i\neq k}\mathbf{Q}_i \mathbf{H}_k^{H} \right)^{-1}$, which entail complexity $\mathcal{O}(M^3)$. However, for $M\geq 3$, we can resort to the matrix inversion lemma to reduce the computational complexity as in \cite{Mccf}, \cite{TMCb}.
	
	Given $\mathbf{t}$ and $\left\{\mathbf{Q}_i\right\}$, we define  $\mathbf{W}_{k,+}\left(\tilde{\mathbf{r}}_{k}^{\hat{\ell}-1,m}\right)=\mathbf{H}_{k} \left(\sum_{i=1}^K\mathbf{Q}_i\right)^{\frac{1}{2}}$, $\mathbf{W}_{k,-}\left(\tilde{\mathbf{r}}_{k}^{\hat{\ell}-1,m}\right)=\mathbf{H}_{k} \left(\sum_{i\neq k}\mathbf{Q}_i\right)^{\frac{1}{2}}$ and denote the $m$-th column vector of $\mathbf{W}_{k,+}^H\left(\tilde{\mathbf{r}}_{k}^{\hat{\ell}-1,m}\right)$ and $\mathbf{W}_{k,-}^H\left(\tilde{\mathbf{r}}_{k}^{\hat{\ell}-1,m}\right)$ by $\mathbf{w}_{k,+}\left(\mathbf{r}_{k,m}^{\hat{\ell}-1}\right)$ and $\mathbf{w}_{k,-}\left(\mathbf{r}_{k,m}^{\hat{\ell}-1}\right)$, respectively, which depend only on the position of receive MA $m$ of UT $k$ and can be written as
	\begin{equation}
		\begin{aligned}
			\mathbf{w}_{k,+}\left(\mathbf{r}_{k,m}^{\hat{\ell}-1}\right)&=\left(\sum_{i=1}^K\mathbf{Q}_i\right)^{\frac{H}{2}}\mathbf{G}_{k}^H\left(\mathbf{t}\right)\mathbf{\Sigma}_k^H \mathbf{f}_{k}\left(\mathbf{r}_{k,m}^{\hat{\ell}-1}\right),\\
			\mathbf{w}_{k,-}\left(\mathbf{r}_{k,m}^{\hat{\ell}-1}\right)&=\left(\sum_{i\neq k}\mathbf{Q}_i\right)^{\frac{H}{2}}\mathbf{G}_{k}^H\left(\mathbf{t}\right)\mathbf{\Sigma}_k^H \mathbf{f}_{k}\left(\mathbf{r}_{k,m}^{\hat{\ell}-1}\right).
		\end{aligned}
	\end{equation}
	Following similar steps as in \cite{Mccf}, we can reformulate the achievable rate of UT $k$ as shown in \eqref{eq: reformulate Rk} at the top of the next page, 
	where $\mathbf{A}_{k,m,+}^{\hat{\ell}}$ and $\mathbf{A}_{k,m,-}^{\hat{\ell}}$ are in fact summations of $M$ rank-one matrices.
	This special structure of MA-enabled MIMO channels can be exploited to promote effective matrix inversions \cite{Mccf}.
	Now, the gradients can be calculated according to \eqref{eq: gradient rkm} at the top of next page,
	\begin{figure*}[t]
		\begin{equation}
			\label{eq: reformulate Rk}
			\begin{aligned}
				&R_k\left(\mathbf{t},\tilde{\mathbf{r}}_{k}^{\hat{\ell}-1,m},\left\{\mathbf{Q}_i\right\},\mathbf{H}_k\right) \\
				=& \log _2 \operatorname{det} \left( \sigma^2\mathbf{I}_{M}+\mathbf{W}_{k,+}\left(\tilde{\mathbf{r}}_{k}^{\hat{\ell}-1,m}\right)\mathbf{W}_{k,+}^H\left(\tilde{\mathbf{r}}_{k}^{\hat{\ell}-1,m}\right)\right) -\log _2 \operatorname{det} \left( \sigma^2\mathbf{I}_{M}+\mathbf{W}_{k,-}\left(\tilde{\mathbf{r}}_{k}^{\hat{\ell}-1,m}\right)\mathbf{W}_{k,-}^H\left(\tilde{\mathbf{r}}_{k}^{\hat{\ell}-1,m}\right)\right)\\
				=& \log _2 \operatorname{det} \underbrace{ \left(	\sigma^2\mathbf{I}_{M}+\mathbf{W}_{k,+}^H\left(\tilde{\mathbf{r}}_{k}^{\hat{\ell}-1,m}\right)\mathbf{W}_{k,+}\left(\tilde{\mathbf{r}}_{k}^{\hat{\ell}-1,m}\right)\right) }_{\triangleq \mathbf{A}_{k,m,+}^{\hat{\ell}}} -\log _2 \operatorname{det} \underbrace{ \left( \sigma^2\mathbf{I}_{M}+\mathbf{W}_{k,-}^H\left(\tilde{\mathbf{r}}_{k}^{\hat{\ell}-1,m}\right)\mathbf{W}_{k,-}\left(\tilde{\mathbf{r}}_{k}^{\hat{\ell}-1,m}\right)\right) }_{\triangleq \mathbf{A}_{k,m,-}^{\hat{\ell}}},\\
				&\mathbf{A}_{k,m,+}^{\hat{\ell}}=	\sigma^2\mathbf{I}_{M}+\sum\nolimits_{m'=1}^{m-1}\mathbf{w}_{k,+}\left(\mathbf{r}_{k,m'}^{\hat{\ell}}\right)\mathbf{w}_{k,+}^H\left(\mathbf{r}_{k,m'}^{\hat{\ell}}\right)+\sum\nolimits_{m'=m}^{M}\mathbf{w}_{k,+}\left(\mathbf{r}_{k,m'}^{\hat{\ell}-1}\right)\mathbf{w}_{k,+}^H\left(\mathbf{r}_{k,m'}^{\hat{\ell}-1}\right),\\
				&\mathbf{A}_{k,m,-}^{\hat{\ell}}=	\sigma^2\mathbf{I}_{M}+\sum\nolimits_{m'=1}^{m-1}\mathbf{w}_{k,-}\left(\mathbf{r}_{k,m'}^{\hat{\ell}}\right)\mathbf{w}_{k,-}^H\left(\mathbf{r}_{k,m'}^{\hat{\ell}}\right)+\sum\nolimits_{m'=m}^{M}\mathbf{w}_{k,-}\left(\mathbf{r}_{k,m'}^{\hat{\ell}-1}\right)\mathbf{w}_{k,-}^H\left(\mathbf{r}_{k,m'}^{\hat{\ell}-1}\right).
			\end{aligned}
		\end{equation}
		\begin{equation}
			\label{eq: gradient rkm}
			\begin{aligned}
				&\nabla_{\mathbf{r}_{k,m}} R_k\left(\mathbf{t}, 	\tilde{\mathbf{r}}_{k}^{\hat{\ell}-1,m},\left\{\mathbf{Q}_i\right\},\mathbf{H}_k\right) =\operatorname{Re}\left\{
				\begin{pmatrix}
					\left( \mathbf{b}_{k,m,+}^{\hat{\ell}} - 	\mathbf{b}_{k,m,-}^{\hat{\ell}} \right) \mathbf{\Delta}_{x_r,k}\mathbf{\Sigma}_k^H \mathbf{f}_{k}\left(\mathbf{r}_{k,m}^{\hat{\ell}-1}\right);
					\left( \mathbf{b}_{k,m,+}^{\hat{\ell}} - 	\mathbf{b}_{k,m,-}^{\hat{\ell}} \right) \mathbf{\Delta}_{y_r,k}\mathbf{\Sigma}_k^H \mathbf{f}_{k}\left(\mathbf{r}_{k,m}^{\hat{\ell}-1}\right)
				\end{pmatrix}\right\},\\
				&\mathbf{b}_{k,m,+}^{\hat{\ell}} \triangleq \frac{2}{\ln 2} 	\mathbf{w}_{k,+}^H\left(\mathbf{r}_{k,m}^{\hat{\ell}-1}\right) \left( \mathbf{A}_{k,m,+}^{\hat{\ell}} \right)^{-1} \left(\sum_{i=1}^K\mathbf{Q}_i\right)^{\frac{H}{2}}\mathbf{G}_{k}^H\left(\mathbf{t}\right),
				\mathbf{b}_{k,m,-}^{\hat{\ell}} \triangleq \frac{2}{\ln 2} 	\mathbf{w}_{k,-}^H\left(\mathbf{r}_{k,m}^{\hat{\ell}-1}\right)\left( \mathbf{A}_{k,m,-}^{\hat{\ell}} \right)^{-1} \left(\sum_{i\neq k}\mathbf{Q}_i\right)^{\frac{H}{2}}\mathbf{G}_{k}^H\left(\mathbf{t}\right).
			\end{aligned}
		\end{equation}
		\hrule
	\end{figure*}
	where matrices $\mathbf{\Delta}_{x_r,k}$ and $\mathbf{\Delta}_{y_r,k}$ are defined in \eqref{eq: Delta} in the Appendix \ref{Ap: gradient rate}.
	
	During the iterations of the GA algorithm, we need to compute a series of matrix inversions in sequence for calculating gradients $\nabla_{\mathbf{r}_{k,m}} R_k$: $\left( \mathbf{A}_{k,1,+}^{1} \right)^{-1}$, $\left( \mathbf{A}_{k,1,-}^{1} \right)^{-1}$, $\cdots$,$\left( \mathbf{A}_{k,M,+}^{\hat{\ell}-1} \right)^{-1}$,$\left( \mathbf{A}_{k,M,-}^{\hat{\ell}-1} \right)^{-1}$,$\left( \mathbf{A}_{k,1,+}^{\hat{\ell}} \right)^{-1}$,$\left( \mathbf{A}_{k,1,-}^{\hat{\ell}} \right)^{-1}$,$\cdots$,$\\\left( \mathbf{A}_{k,M,+}^{\hat{\ell}}\right)^{-1}$,$\left( \mathbf{A}_{k,M,-}^{\hat{\ell}}\right)^{-1}$, $\cdots$. To reduce the computational complexity, matrix inversions can be obtained based on the previous one. That is, we first compute inverse matrices $\left( \mathbf{A}_{k,1,+}^1 \right)^{-1}$ and $\left( \mathbf{A}_{k,1,-}^1 \right)^{-1}$ and then 
	update $\left( \mathbf{A}_{k,m,+}^{\hat{\ell}} \right)^{-1}$ and $\left( \mathbf{A}_{k,m,-}^{\hat{\ell}} \right)^{-1}$ based on $\left( \mathbf{A}_{k,m',+}^{\hat{\ell}'} \right)^{-1}$ and $\left( \mathbf{A}_{k,m',-}^{\hat{\ell}'} \right)^{-1}$, respectively, where $m'=m-1$, $\hat{\ell}'=\hat{\ell}$ when $m=2,3,\cdots,M$ and $m'=M$, $\hat{\ell}'=\hat{\ell}-1$ when $m=1$. Specifically, we define $\mathbf{Z}_{1,+}=\left[\mathbf{w}_{k,+}\left(\mathbf{r}_{k,m'}^{\hat{\ell}'}\right), \mathbf{w}_{k,+}\left(\mathbf{r}_{k,m'}^{\hat{\ell}'-1}\right)\right] \in \mathbb{C}^{N \times 2}$ and $\mathbf{Z}_{2,+}=\left[\mathbf{w}_{k,+}\left(\mathbf{r}_{k,m'}^{\hat{\ell}'}\right),-\mathbf{w}_{k,+}\left(\mathbf{r}_{k,m'}^{\hat{\ell}'-1}\right)\right] \in \mathbb{C}^{N \times 2}$ such that
	\begin{equation}
		\mathbf{A}_{k,m,+}^{\hat{\ell}} = \mathbf{A}_{k,m',+}^{\hat{\ell}'}+\mathbf{Z}_{1,+} \mathbf{Z}_{2,+}^H.
	\end{equation}
	According to the matrix inversion lemma, we have \cite{TMCb}
	\begin{equation}
		\begin{aligned}
			&\left(\mathbf{A}_{k,m,+}^{\hat{\ell}}\right)^{-1}=\left(\mathbf{A}_{k,m',+}^{\hat{\ell}'}\right)^{-1}-\left(\mathbf{A}_{k,m',+}^{\hat{\ell}'}\right)^{-1} \mathbf{Z}_{1,+}\\&\times \left(\mathbf{I}_2+\mathbf{Z}_{2,+}^H \left(\mathbf{A}_{k,m',+}^{\hat{\ell}'}\right)^{-1} \mathbf{Z}_{1,+}\right)^{-1} \mathbf{Z}_{2,+}^H \left(\mathbf{A}_{k,m',+}^{\hat{\ell}'}\right)^{-1},
		\end{aligned}
	\end{equation}
	where only a 2-by-2 matrix inversion needs to be calculated.
	Similarly, the inverse matrix of $\mathbf{A}_{k,m,-}^{\hat{\ell}}$ can be calculated as
	\begin{equation}
		\begin{aligned}
			&\left(\mathbf{A}_{k,m,-}^{\hat{\ell}}\right)^{-1}=\left(\mathbf{A}_{k,m',-}^{\hat{\ell}'}\right)^{-1}-\left(\mathbf{A}_{k,m',-}^{\hat{\ell}'}\right)^{-1} \mathbf{Z}_{1,-}\\&\times \left(\mathbf{I}_2+\mathbf{Z}_{2,-}^H \left(\mathbf{A}_{k,m',-}^{\hat{\ell}'}\right)^{-1} \mathbf{Z}_{1,-}\right)^{-1} \mathbf{Z}_{2,-}^H \left(\mathbf{A}_{k,m',-}^{\hat{\ell}'}\right)^{-1},
		\end{aligned}
	\end{equation}
	where $\mathbf{Z}_{1,-}\triangleq\left[\mathbf{w}_{k,-}\left(\mathbf{r}_{k,m'}^{\hat{\ell}'}\right), \mathbf{w}_{k,-}\left(\mathbf{r}_{k,m'}^{\hat{\ell}'-1}\right)\right]$ and $\mathbf{Z}_{2,-}\triangleq\left[\mathbf{w}_{k,-}\left(\mathbf{r}_{k,m'}^{\hat{\ell}'}\right),-\mathbf{w}_{k,-}\left(\mathbf{r}_{k,m'}^{\hat{\ell}'-1}\right)\right]$.
	
	\begin{algorithm}
		\caption{GA Algorithm for Solving Problem \eqref{op: decouple S}}
		\label{Al: Rx r GA}
		\begin{spacing}{1}
			\textbf{Input:} $\mathbf{t}$, $\tilde{\mathbf{r}}_k^0$ and $\left\{\mathbf{Q}_i\right\}$; $M$, $\sigma^2$, $\lambda$, $\mathcal{C}_{r,k}$; $s$, $\tau$, $\xi$, $\epsilon$; I-CSIR \\
			\textbf{Output:} $\mathbf{r}_{k}^{\ell,b}$
			\begin{algorithmic}[1]
				\STATE {Initialize $\tilde{\mathbf{r}}_k^0$ satisfying the position constraints and $\hat{\ell}=1$}
				\REPEAT
				\FOR{$m=1,2,\cdots,M$}
				\STATE{Calculate $\nabla_{\mathbf{r}_{k,m}} R_k\left(\mathbf{t}, \tilde{\mathbf{r}}_{k}^{\hat{\ell}-1,m},\left\{\mathbf{Q}_i\right\},\mathbf{H}_k\right)$ by \eqref{eq: gradient rkm}} 
				\STATE{Set $\tau^{\hat{\ell},m}=s$ and update $\mathbf{r}_{k,m}^{\hat{\ell}}$ by \eqref{eq: RMA update}}
				\WHILE{\eqref{eq: Armijo–Goldstein condition} and $\left\|\mathbf{r}_{k,i}-\mathbf{r}_{k,m}\right\|^2 \geq D^2$, $\forall i\neq m$ are not satisfied}
				\STATE{Shrink the step size $\tau^{\hat{\ell},m}=\tau\tau^{\hat{\ell},m}$}
				\STATE{Update $\mathbf{r}_{k,m}^{\hat{\ell}}$ by \eqref{eq: RMA update}}
				\ENDWHILE
				\ENDFOR
				\STATE{$\hat{\ell}=\hat{\ell}+1$}
				\UNTIL{the increment of $R_k\left(\mathbf{t}, \tilde{\mathbf{r}}_{k},\left\{\mathbf{Q}_i\right\},\mathbf{H}_k\right)$ is less than $\epsilon$\\}
				\STATE{Set $\mathbf{r}_{k}^{\ell,b}=\tilde{\mathbf{r}}_{k}^{\hat{\ell},1}$}
				\RETURN{$\mathbf{r}_{k}^{\ell,b}$}
			\end{algorithmic}
		\end{spacing}
	\end{algorithm}
	The overall solution of problem \eqref{op: decouple S} is summarized in Algorithm \ref{Al: Rx r GA}. In Step 1, the receive APV is initialized as $\tilde{\mathbf{r}}_k^0$ to satisfy the position constraints. Subsequently, the receive APVs are optimized in Steps 2-12. The convergence of Algorithm \ref{Al: Rx r GA} is analyzed as follows. Since $R_k\left(\mathbf{t},\tilde{\mathbf{r}}_k,\left\{\mathbf{Q}_i\right\},\mathbf{H}_k\right)$ and C3$_k$ are continuous functions w.r.t. $\mathbf{r}_{k,m}$, if one of the elements in $\nabla_{\mathbf{r}_{k,m}} R_k\left(\mathbf{t}, \tilde{\mathbf{r}}_{k}^{\hat{\ell}-1,m},\left\{\mathbf{Q}_i\right\},\mathbf{H}_k\right)$ is not equal to zero and the corresponding gradient direction points towards the inside of the feasible region, we can always find a sufficiently small positive $\tau^{\hat{\ell},m}$ to satisfy the Armijo–Goldstein condition in \eqref{eq: Armijo–Goldstein condition} and the antenna distance constraints, i.e., $\left\|\mathbf{r}_{k,i}-\mathbf{r}_{k,m}\right\|^2 \geq D^2$, $\forall i\neq m$. Thus, in each iteration, the update of the position of each MA in Steps 4-9 ensure that the objective value is non-decreasing. Besides, $R_k\left(\mathbf{t},\tilde{\mathbf{r}}_k,\left\{\mathbf{Q}_i\right\},\mathbf{H}_k\right)$ is upper-bounded by a finite value since the feasible region is compact. Therefore, we can conclude that the convergence of Algorithm \ref{Al: Rx r GA} is guaranteed. The computational complexity of Algorithm \ref{Al: Rx r GA} is mainly determined by the required matrix inversions, which entail a complexity of $\mathcal{O}(M)$ in each outer iteration.
	As a result, the computational complexity of Algorithm \ref{Al: Rx r GA} is $\mathcal{O}(\hat{I}M)$, where $\hat{I}$ is the maximum number of outer iterations. 
	
\subsection{Long-term Transmit Design Based on S-CSIT}
The main idea behind the CSSCA algorithm is to iteratively optimize a sequence of deterministic convex objective/feasibility optimization problems obtained by replacing the objective/constraint functions in the original complicated problem with tractable convex surrogate functions \cite{SSCA}. On the other hand, when multiple system state samples are exploited in each iteration, the surrogate functions constructed are more accurate and thus the number of iterations required for convergence can potentially be reduced, while the complexity per iteration increases. Therefore, we leverage multiple samples of the system state in each iteration, which can efficiently balance the tradeoff between the required number of iterations and the complexity per iteration \cite{SSCA,TSSO}.

Then, in iteration $\ell$ of the proposed iterative CSSCA algorithm, $B$ realization of the CSI, i.e., $\mathbf{H}_k^{\ell,b}$, $\forall k$, $\forall b$, are generated based on the distribution of the S-CSIT. Then, the average achievable rate of UT $k$, i.e., the left hand-side term in C1$'_k$, is replaced by the recursive concave surrogate function
\begin{equation}
	\label{eq: surrogate functions rate}
	\begin{aligned} f_k^{\ell}\left(\mathbf{t},\left\{\mathbf{Q}_i\right\}\right) &= \left(1-\rho^{\ell}\right) f_k^{\ell-1}\left(\mathbf{t},\left\{\mathbf{Q}_i\right\}\right) + \rho^{\ell} g_{k}^{\ell}\left(\mathbf{t},\left\{\mathbf{Q}_i\right\}\right), 
	\end{aligned}
\end{equation}
where the initial values are set as $ f_k^0\left(\mathbf{t},\left\{\mathbf{Q}_i\right\}\right) = 0$. Sequence $\rho^{\ell} \in(0,1]$ is chosen to satisfy $\rho^{\ell} > 0$, $\sum_{\ell=1}^{\infty} \rho^{\ell}=\infty$, $\sum_{\ell=1}^{\infty} (\rho^{\ell})^2 < \infty$. We select sample surrogate function, $ g_{k}^{\ell}\left(\mathbf{t},\left\{\mathbf{Q}_i\right\}\right)$, as concave function given by \cite{SSCA}
\begin{equation}
	\label{eq: sample surrogate functions}
	\begin{aligned}
		&g_{k}^{\ell}\left(\mathbf{t},\left\{\mathbf{Q}_i\right\}\right) = R_k^{\ell} + \left( \mathbf{\Delta}_{\mathbf{t},k}^{\ell} \right)^T \left(\mathbf{t}-\mathbf{t}^{\ell}\right)+ \tau_{t,k} \left\|\mathbf{t}-\mathbf{t}^{\ell}\right\|^2 \\
		&+ \sum_{i=1}^K \operatorname{tr}\left( \left( \mathbf{\Delta}_{\mathbf{Q}_i,k}^{\ell} \right)^H \left(\mathbf{Q}_i-\mathbf{Q}_i^{\ell}\right)\right) + \tau_{Q,k,i} \left\|\mathbf{Q}_i-\mathbf{Q}_i^{\ell}\right\|^2,
	\end{aligned}
\end{equation}
where coefficients $\left\{\tau_{t,k}\right\}$ and $\left\{\tau_{Q,k,i}\right\}$ are chosen to be negative constants to ensure that the surrogate functions are uniformly strongly convex in $\mathbf{t}$ and $\left\{\mathbf{Q}_i\right\}$.
The approximate gradients $\mathbf{\Delta}_{\mathbf{t},k}^{\ell}$ and $\mathbf{\Delta}_{\mathbf{Q}_i,k}^{\ell}$ as well as function value $R_k^{\ell}$ are given by
\begin{subequations}
	\label{eq: gradients}
	\begin{align}
		\mathbf{\Delta}_{\mathbf{t},k}^{\ell} &\triangleq \frac{1}{B} \sum\nolimits_{b=1}^B \nabla_{\mathbf{t}} R_k\left(\mathbf{t}^{\ell}, \mathbf{r}_k^{\ell,b},\left\{\mathbf{Q}_i^{\ell}\right\},\mathbf{H}_k^{\ell,b}\right), \\
		\mathbf{\Delta}_{\mathbf{Q}_i,k}^{\ell} &\triangleq \frac{1}{B} \sum\nolimits_{b=1}^B \nabla_{\mathbf{Q}_i} R_k\left(\mathbf{t}^{\ell}, \mathbf{r}_k^{\ell,b},\left\{\mathbf{Q}_i^{\ell}\right\},\mathbf{H}_k^{\ell,b}\right), \\
		R_k^{\ell} &\triangleq \frac{1}{B} \sum\nolimits_{b=1}^B R_k\left(\mathbf{t}^{\ell}, \mathbf{r}_k^{\ell,b},\left\{\mathbf{Q}_i^{\ell}\right\},\mathbf{H}_k^{\ell,b}\right),
	\end{align}
\end{subequations}
respectively, where $\mathbf{r}_k^{\ell,b}$ is the suboptimal solution obtained from short-term problem \eqref{op: decouple S} for $\mathbf{H}_k^{\ell,b}$ and the gradients of the achievable rate, $\nabla_{\mathbf{t}} R_k\left(\mathbf{t}^{\ell}, \mathbf{r}_k^{\ell,b},\left\{\mathbf{Q}_i^{\ell}\right\},\mathbf{H}_k^{\ell,b}\right)$ and $ \nabla_{\mathbf{Q}_i} R_k\left(\mathbf{t}^{\ell}, \mathbf{r}_k^{\ell,b},\left\{\mathbf{Q}_i^{\ell}\right\},\mathbf{H}_k^{\ell,b}\right)$, are given in Appendix \ref{Ap: gradient rate}. Moreover, for the left hand-side term of constraint C2, which does not involve random system states, we adopt the following concave surrogate function
\begin{equation}
	\label{eq: surrogate functions constraints}
	\begin{aligned}
		h_{i,j}^{\ell}\left(\mathbf{t}_i,\mathbf{t}_j\right) &=\tau_{h_{i,j}}\left(\left\|\mathbf{t}_i-\mathbf{t}_i^{\ell}\right\|^2
		+\left\|\mathbf{t}_j-\mathbf{t}_j^{\ell}\right\|^2\right)\\&+2\left(\mathbf{t}_i^{\ell}-\mathbf{t}_j^{\ell}\right)^{\operatorname{T}} \left(\mathbf{t}_i-\mathbf{t}_j\right)
		-\left\|\mathbf{t}_i^{\ell}-\mathbf{t}_j^{\ell}\right\|^2.
	\end{aligned}
\end{equation}
The coefficients of the surrogate function in \eqref{eq: surrogate functions constraints}, $\left\{\tau_{h_{i,j}}\right\}$, are chosen as negative constants to ensure concavity.

According to the CSSCA algorithm framework for solving problem \eqref{op: decouple L} \cite{SSCA}, the optimal solution to the following problem in iteration $\ell$, $\bar{\mathbf{t}}^{\ell}$ and $\left\{\bar{\mathbf{Q}}_i^{\ell}\right\}$, is obtained (if it is feasible) as:
	\begin{equation}
		\label{op: surrogate 1}
		\begin{aligned}
			\underset{\mathbf{t}\in \mathcal{C}_t, \left\{\mathbf{Q}_i\succeq\mathbf 0\right\}} {\operatorname{max}}  \quad &\sum_{k=1}^K f_{k}^{\ell}\left(\mathbf{t},\left\{\mathbf{Q}_i\right\}\right) \\
			\text {s.t. } \qquad
			&\text{C1a$'_k$: } f_{k}^{\ell}\left(\mathbf{t},\left\{\mathbf{Q}_i\right\}\right) \geq R_{\mathrm{min}}, \,\, \forall k, \\
			&\text{C2$'$: } h_{i,j}^{\ell}\left(\mathbf{t}_i,\mathbf{t}_j\right) \geq D^2, \forall i \neq j,\\
			&\text{C4},
		\end{aligned}
	\end{equation}
	which is a concave approximation of problem \eqref{op: decouple L}.
	If problem \eqref{op: surrogate 1} is not feasible, i.e., constraints C1a$'_k$, $\forall k$, cannot be satisfied, the following feasibility problem is solved instead \cite{SSCA}:
	\begin{equation}
		\label{op: surrogate 2}
		\begin{aligned}
			\underset{\mathbf{t}\in \mathcal{C}_t, \left\{\mathbf{Q}_i\succeq\mathbf 0\right\}, \alpha} {\operatorname{max}} \quad&\quad \alpha \\
			\text {s.t. } \,\,\qquad &
			\text{C1b$'_k$: } f_{k}^{\ell}\left(\mathbf{t},\left\{\mathbf{Q}_i\right\}\right) \geq R_{\mathrm{min}}+\alpha, \,\, \forall k, \\
			&\text{C2$'$}, \text{C4}, 
		\end{aligned}
	\end{equation}
	which aims to maximize the minimum rate to effectively minimize the violation of constraints C1a$'_k$, $\forall k$. After the optimal solution to the approximate objective/feasibility problem, $\bar{\mathbf{t}}^{\ell}$ and $\left\{\bar{\mathbf{Q}}_i^{\ell}\right\}$, has been obtained, the optimization variables are updated as follows \cite{SSCA}
	\begin{subequations}
		\label{eq: update variables}
		\begin{align}
			\mathbf{t}^{\ell+1}&=\left(1-\gamma^{\ell}\right)\mathbf{t}^{\ell}+\gamma^{\ell}\bar{\mathbf{t}}^{\ell}, \label{eq: update t}\\
			\mathbf{Q}_k^{\ell+1}&=\left(1-\gamma^{\ell}\right)\mathbf{Q}_k^{\ell}+\gamma^{\ell}\bar{\mathbf{Q}}_k^{\ell}, \forall k, \label{eq: update Q}
		\end{align}
	\end{subequations}
	where $\gamma^{\ell} \in(0,1]$ is a decreasing sequence satisfying $\gamma^{\ell} \rightarrow 0$, $\sum_{\ell=1}^{\infty} \gamma^{\ell}=\infty$, $ \sum_{\ell=1}^{\infty}\left(\gamma^{\ell}\right)^2<\infty$, $\lim_{\ell \rightarrow \infty} \gamma^{\ell} /\rho^{\ell}=0$. Note that we can always find a point, $\mathbf{t}$, satisfying constraints C2$'$ if the initial point, $\mathbf{t}^1$, is chosen to be feasible since the following inequalities holds:
	\begin{equation}
		\begin{aligned}
			\left\|\mathbf{t}_i^{\ell+1}-\mathbf{t}_j^{\ell+1}\right\|^2 
			\overset{\text{(a)}}{\geq}& h_{i,j}^{\ell}\left(\left(1-\gamma^{\ell}\right)\mathbf{t}_i^{\ell}+\gamma^{\ell}\bar{\mathbf{t}}_i^{\ell},\left(1-\gamma^{\ell}\right)\mathbf{t}_j^{\ell}+\gamma^{\ell}\bar{\mathbf{t}}_j^{\ell}\right)
			\\\overset{\text{(b)}}{\geq}& \left(1-\gamma^{\ell}\right)h_{i,j}^{\ell}\left(\mathbf{t}_i^{\ell},\mathbf{t}_j^{\ell}\right) + \gamma^{\ell} h_{i,j}^{\ell}\left(\bar{\mathbf{t}}_i^{\ell}, \bar{\mathbf{t}}_j^{\ell}\right)
			\\=& \left(1-\gamma^{\ell}\right)\left\|\mathbf{t}_i^{\ell}-\mathbf{t}_j^{\ell}\right\|^2 + \gamma^{\ell} h_{i,j}^{\ell}\left(\bar{\mathbf{t}}_i^{\ell}, \bar{\mathbf{t}}_j^{\ell}\right) \\\geq& D^2.
		\end{aligned}
	\end{equation}
	Here, (a) holds due to the fact that $\left\|\mathbf{t}_i-\mathbf{t}_j\right\|^2$ is a convex function w.r.t. $\mathbf{t}_i$ and $\mathbf{t}_j$, i.e., $\nabla^2\left\|\mathbf{t}_i-\mathbf{t}_j\right\|^2\succeq \mathbf{0}$.
	Thus, based on Taylor's theorem, $\left\|\mathbf{t}_i-\mathbf{t}_j\right\|^2$ is globally lower bounded by the concave surrogate function in \eqref{eq: surrogate functions constraints} \cite{MDCW}. Besides, inequality (b) holds due to the concavity of the surrogate function in \eqref{eq: surrogate functions constraints}.

	\begin{algorithm}[t]
		\caption{CSSCA Algorithm for Solving Problem \eqref{op: decouple L}}
		\label{Al: CSSCA t r Q}
		\begin{spacing}{1}
			\textbf{Input:} $N$, $M$, $K$, $\sigma^2$, $\lambda$, $\mathcal{C}_{t}$, $\{\mathcal{C}_{r,k}\}$; $\left\{\tau_{t,k}\right\}$, $\left\{\tau_{r,k}\right\}$, $\left\{\tau_{Q,k,i}\right\}$, $\left\{\tau_{h_{i,j}}\right\}$, $I$; S-CSIT \\
			\textbf{Output:} $\mathbf{t}^*$, $\left\{\mathbf{Q}_i^*\right\}$
			\begin{algorithmic}[1]
				\STATE {Initialize $\mathbf{t}^1$ and $\left\{\mathbf{Q}_i^1\right\}$ satisfying the position and power constraints, respectively}
				\FOR{$\ell = 1 \text{ to } I$}
				\STATE {Obtain a mini-batch $\left\{ \mathbf{H}_k^{\ell,b}, b=1,\cdots,B \right\}$ and then compute $\left\{\mathbf{r}_k^{\ell,b}\right\}$ by Algorithm \ref{Al: Rx r GA}}
				\STATE {Construct the surrogate functions $ f_k^{\ell}\left(\mathbf{t},\left\{\mathbf{Q}_i\right\}\right)$, $\forall k$, according to \eqref{eq: surrogate functions rate}}
				\IF{problem \eqref{op: surrogate 1} is feasible}
				\STATE {Obtain $\bar{\mathbf{t}}^{\ell}$ and $\left\{\bar{\mathbf{Q}}_i^{\ell}\right\}$ by solving \eqref{op: surrogate 1}}
				\ELSE
				\STATE {Obtain $\bar{\mathbf{t}}^{\ell}$ and $\left\{\bar{\mathbf{Q}}_i^{\ell}\right\}$ by solving \eqref{op: surrogate 2}}
				\ENDIF
				\STATE{Update $\mathbf{t}^{\ell+1}$ and $\left\{\mathbf{Q}_i^{\ell+1}\right\}$ according to \eqref{eq: update variables}}
				\ENDFOR
				\STATE {Set $\mathbf{t}^*=\mathbf{t}^{I+1}$, $\left\{\mathbf{Q}_i^*=\mathbf{Q}^{I+1}_i\right\}$}
				\STATE{\textbf{return} $\mathbf{t}^*$, $\left\{\mathbf{Q}_i^*\right\}$}
			\end{algorithmic}
		\end{spacing}
	\end{algorithm}
	The CSSCA-based solution for solving problem \eqref{op: decouple L} is summarized in Algorithm \ref{Al: CSSCA t r Q}, where $I$ denotes the total number of iterations. In Step 1, the APV $\mathbf{t}^1$ and $\left\{\mathbf{Q}_i^1\right\}$ are initialized to satisfy the position and power constraints, respectively. In Steps 2-11, the transmit APV and covariance matrices are jointly optimized. 
	The computational complexity of Algorithm \ref{Al: CSSCA t r Q} is analyzed as follows. Firstly, obtaining a series of suboptimal solution to problem \eqref{op: decouple S}, $\left\{\mathbf{r}_k^{\ell,b}\right\}$, entails complexity $\mathcal{O}\left(BK\hat{I}M\right)$. Secondly, the complexities of calculating $\nabla_{\mathbf{t}} R_k\left(\mathbf{t}^{\ell}, \mathbf{r}_k^{\ell,b},\left\{\mathbf{Q}_i^{\ell}\right\},\mathbf{H}_k^{\ell,b}\right)$ and $ \nabla_{\mathbf{Q}_i} R_k\left(\mathbf{t}^{\ell}, \mathbf{r}_k^{\ell,b},\left\{\mathbf{Q}_i^{\ell}\right\},\mathbf{H}_k^{\ell,b}\right)$, $i=1,2,\cdots,K$, $\forall k$, are $\mathcal{O}\left(K M^{3}\right)$, respectively. Finally, the complexity of solving the objective/feasibility problem in each iteration of Algorithm \ref{Al: CSSCA t r Q} by the interior-point method with accuracy $\epsilon$ is $\mathcal{O}\left(\log(1/\epsilon)N^7\right)$ \cite{CO}. Thus, the computational complexity of Algorithm \ref{Al: CSSCA t r Q} is $\mathcal{O}\left(I\left(BK\hat{I}M+K M^{3}+\log(1/\epsilon)N^7\right)\right)$.
	
	Note that the GMM of the MAs considered so far allows unrestricted movement of each transmit and receive MA within the given regions (i.e., permitting any two dimensional displacement). However, the large repositioning distances may lead to high power consumption and long delays, which are not desirable for the receive MAs as they are frequently updated based on the I-CSIR. Conversely, since S-CSIT tends to remain constant over a much longer period, large repositioning distances are less critical for the transmit MAs. As a result,  in the subsequent section, we propose a PMM for the receive MAs, while still employing the GMM for the transmit MAs.
	
	\section{Antenna Movement Mode Design}
	\label{section: Antenna Move Mode Design}
\begin{figure}
		\centering
		\includegraphics[width=0.49\textwidth,height=0.3\textwidth]{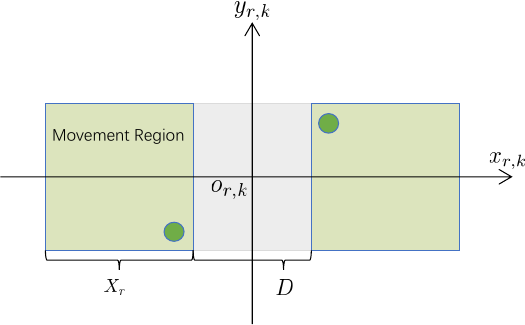}
		\caption{The proposed PMM architecture (circles represent receive MAs).}
		\label{fig: MA Planar}
	\end{figure}

	As illustrated in Fig.~\ref{fig: MA Planar}, each receive MA is only allowed to move in a given planar region, which does not overlap with the regions of the other MAs. The movement region for the $m$-th receive MA of UT $k$ is denoted as $\mathcal{C}_{r,k,m}$ of size $X_r\times X_r$.
	Moreover, the minimum distance between any two regions is set to $D$ to avoid potential coupling effects.  With a slight abuse of notation, we redefine $\mathcal{C}_{r,k} \triangleq \mathcal{C}_{r,k,1}\times \cdots \times \mathcal{C}_{r,k,M}$. 
	
	\subsection{PDD-SSCA Framework for PMM}
	Adopting the PMM, the problem in \eqref{op: original 1} can be written as
	\begin{equation}
		\label{op: original 2}
		\begin{aligned}
			\max _{\mathbf{t}\in \mathcal{C}_t, \left\{\mathbf{Q}_i\succeq\mathbf 0\right\}} \quad 	&\sum_{k=1}^K \bar R_k\left(\mathbf{t},\left\{\mathbf{Q}_i\right\}\right) \\
			\text { s.t. } \qquad
			&\text{C1$_k$}, \forall k,\text{C2},\text{C4}.
		\end{aligned}
	\end{equation}
	Since constraint $\tilde{\mathbf{r}}_k\in \mathcal{C}_{r,k}$ in \eqref{eq: Rbar} is convex for the PMM, an efficient PDD-SSCA algorithm can be developed to solve problem \eqref{op: original 2}, where a novel primal-dual decomposition method is employed to decouple the two-timescale optimization variables \cite{TSSO}. 
	
	Specifically, we decompose problem \eqref{op: original 2} into one long-term problem and $K$ short-term problems (each corresponding to one channel sample $\mathbf{H}_k$). Each receive APV, which does not involve coupled constraints between MAs thanks to the PMM, is updated as a whole efficiently by the GP method \cite{CO,Np} to obtain a stationary point of the corresponding short-term problem, while the long-term problem is still tackled using the CSSCA algorithm framework \cite{SSCA}. Note that the dual variables of the PDD-SSCA algorithm framework can be omitted here, due to the structure of problem \eqref{op: original 2}.
	
	Given long-term variables $\mathbf{t}$ and $\left\{\mathbf{Q}_i\right\}$, let $\tilde{\mathbf{r}}_k^{\hat{I}}\left(\mathbf{t},\left\{\mathbf{Q}_i\right\},\mathbf{H}_k\right)$ denote a stationary point obtained by running the GP algorithm for a sufficiently large number of iterations, $\hat{I}$, for the following short-term problem for UT $k$:
	\begin{equation}
		\label{op: PDD S}
		\begin{aligned}
			\max _{\tilde{\mathbf{r}}_k\in \mathcal{C}_{r,k}} \quad 	& R_k\left(\mathbf{t},\tilde{\mathbf{r}}_k,\left\{\mathbf{Q}_i\right\},\mathbf{H}_k\right).
		\end{aligned}
	\end{equation} 
	Then, with the short-term policy $\left\{\tilde{\mathbf{r}}_k^{\hat{I}}\left(\mathbf{t},\left\{\mathbf{Q}_i\right\},\mathbf{H}_k\right)\right\}$, we can formulate the following long-term problem:
	\begin{equation}
		\label{op: PDD L}
		\begin{aligned}
			\max _{\mathbf{t}\in \mathcal{C}_t, \left\{\mathbf{Q}_i\succeq\mathbf 0\right\}} \quad 	&\sum_{k=1}^K \mathbb{E}_{\mathbf{H}_k} \left\{ R_k\left(\mathbf{t},\tilde{\mathbf{r}}_k^{\hat{I}},\left\{\mathbf{Q}_i\right\},\mathbf{H}_k\right) \right\} \\
			\text { s.t. } \qquad
			&\text{C1$''_k$}, \forall k,\text{C2},\text{C4},
		\end{aligned}
	\end{equation}
	where constraint $\mathbb{E}_{\mathbf{H}_k} \left\{ R_k\left(\mathbf{t},\tilde{\mathbf{r}}_k^{\hat{I}},\left\{\mathbf{Q}_i\right\},\mathbf{H}_k\right) \right\} \geq R_{\mathrm{min}}$ is denoted as C1$''_k$.
	
	\subsection{Short-term Receive APV Design Based on I-CSIR}
	\label{subsection: Short-term Receive APV Design with I-CSI}
	For given $\mathbf{t}$ and $\left\{\mathbf{Q}_i\right\}$, problem \eqref{op: PDD S} only contains
	convex constraints while the objective function is still nonconvex. Therefore, it is computationally efficient to iteratively obtain a stationary solution using a GP algorithm \cite{Np}. In particular, in iteration $\hat{\ell}$,  the optimization variables are updated by moving along the gradient direction, i.e.,
	\begin{equation}
		\label{eq: gp update}
		\begin{aligned}
			\tilde{\mathbf{r}}_k^{\hat{\ell}} &= \left[\tilde{\mathbf{r}}_k^{\hat{\ell}-1}+\tau^{\hat{\ell}}\nabla_{\mathbf{r}_k} R_k\left(\mathbf{t},\tilde{\mathbf{r}}_k^{\hat{\ell}-1},\left\{\mathbf{Q}_i\right\},\mathbf{H}_k\right) \right]_{\mathcal{C}_{r,k}},
		\end{aligned}
	\end{equation}
	where $\tau^{\hat{\ell}}$ is a step size properly chosen by the backtracking line search, similar to that in Algorithm \ref{Al: Rx r GA}, such that it satisfies the following Armijo–Goldstein condition: 
	\begin{equation}
		\begin{aligned}
			R_k&\left(\mathbf{t},\tilde{\mathbf{r}}_{k}^{\hat{\ell}},\left\{\mathbf{Q}_i\right\},\mathbf{H}_k \right) \geq R_k\left(\mathbf{t},\tilde{\mathbf{r}}_{k}^{\hat{\ell}-1},\left\{\mathbf{Q}_i\right\},\mathbf{H}_k \right) \\&
			+ \xi \tau^{\hat{\ell}} \left\|\nabla_{\mathbf{r}_{k}} R_k\left(\mathbf{t}, \tilde{\mathbf{r}}_{k}^{\hat{\ell}-1},\left\{\mathbf{Q}_i\right\},\mathbf{H}_k\right)\right\|^2. \label{eq: Armijo–Goldstein condition PMM}
		\end{aligned}
	\end{equation}
	
	Define the binary diagonal matrices $\mathbf{B}_k^{\hat{\ell}}$, $\hat{\ell}=1,\cdots,\hat{I}$, where the $j$-th main diagonal entry $\left[\mathbf{B}_k^{\hat{\ell}}\right]_{j,j}=0$, when $\left[\tilde{\mathbf{r}}_k^{\hat{\ell}-1}+\tau^{\hat{\ell}}\nabla_{\mathbf{r}_k} R_k\left(\mathbf{t},\tilde{\mathbf{r}}_k^{\hat{\ell}-1},\left\{\mathbf{Q}_i\right\},\mathbf{H}_k\right)\right]_j$ is outside of its domain of definition (i.e., the movement region corresponding to the $j$-th dimension), and $\left[\mathbf{B}_k^{\hat{\ell}}\right]_{j,j}=1$, otherwise.
	Thus, $\tilde{\mathbf{r}}_k^{\hat{I}}\left(\mathbf{t},\left\{\mathbf{Q}_i\right\},\mathbf{H}_k\right)$ satisfies the following inequality due to the KKT conditions of problem \eqref{op: PDD S}:
	\begin{equation}
		\label{eq: short-term KKT}
		\begin{aligned}
			\left\| \mathbf{B}_k^{\hat{I}} \nabla_{\mathbf{r}_k} R_k\left(\mathbf{t},\tilde{\mathbf{r}}_k^{\hat{I}},\left\{\mathbf{Q}_i\right\},\mathbf{H}_k\right) \right\|\leq e^{\hat{I}}(\mathbf{t},\left\{\mathbf{Q}_i\right\}),
		\end{aligned}
	\end{equation}
	where $e^{\hat{I}}(\mathbf{t},\left\{\mathbf{Q}_i\right\})$ denotes the error caused because the short-term algorithm only runs for a finite number of $\hat{I}$ iterations. 
	Then, for all $\mathbf{t}$ and $\left\{\mathbf{Q}_i\right\}$, we have $\lim _{\hat{I} \rightarrow \infty} e^{\hat{I}}(\mathbf{t},\left\{\mathbf{Q}_i\right\})=0$.
	
	The GP algorithm for solving problem \eqref{op: PDD S} is summarized in Algorithm \ref{Al: Rx r GP}. $\tilde{\mathbf{r}}_k^{\hat{\ell}}$ is initialized to satisfy the position constraints in Step 1 and then updated in Steps 2-9. Algorithm \ref{Al: Rx r GP} converges to a stationary point when the step size sequence is properly chosen \cite{Np}. The computational complexity of Algorithm \ref{Al: Rx r GP} is mainly determined by the matrix inversions needed for calculating gradients $\nabla_{\mathbf{r}_k} R_k\left(\mathbf{t},\tilde{\mathbf{r}}_k^{\hat{\ell}-1},\left\{\mathbf{Q}_i\right\},\mathbf{H}_k\right)$, which entails complexity $\mathcal{O}\left(M^3\right)$ in each iteration. Therefore, the complexity of Algorithm \ref{Al: Rx r GP} is $\mathcal{O}\left(\hat{I} M^3\right)$, where $\hat{I}$ is the number of iterations.
		
	\begin{algorithm}[t]
		\caption{GP Algorithm for Solving Problem \eqref{op: PDD S}}
		\label{Al: Rx r GP}
		\begin{spacing}{1}
			\textbf{Input:}  $\mathbf{t}$, $\tilde{\mathbf{r}}_k^0$ and $\left\{\mathbf{Q}_i\right\}$; $M$, $\sigma^2$, $\lambda$, $\mathcal{C}_{r,k}$; $\hat{I}$; I-CSIR \\
			\textbf{Output:} $\mathbf{r}_k^{\ell,b}$
			\begin{algorithmic}[1]
				\STATE {Initialize $\tilde{\mathbf{r}}_k^0$ satisfying the position constraints}
				\FOR{$\hat{\ell} = 1 \text{ to } \hat{I}$}
				\STATE{Calculate $\nabla_{\mathbf{r}_{k}} R_k\left(\mathbf{t}, \tilde{\mathbf{r}}_{k}^{\hat{\ell}-1},\left\{\mathbf{Q}_i\right\},\mathbf{H}_k\right)$ by \eqref{eq: rate differential r}} 
				\STATE{Set $\tau^{\hat{\ell}}=s$ and update $\tilde{\mathbf{r}}_k^{\hat{\ell}}$ by \eqref{eq: gp update}}
				\WHILE{\eqref{eq: Armijo–Goldstein condition PMM} is not satisfied}
				\STATE{Shrink the step size $\tau^{\hat{\ell}}=\tau\tau^{\hat{\ell}}$}
				\STATE{Update $\tilde{\mathbf{r}}_k^{\hat{\ell}}$ by \eqref{eq: gp update}}
				\ENDWHILE
				\ENDFOR
				\STATE {Set $\mathbf{r}_k^{\ell,b}=\tilde{\mathbf{r}}_k^{\hat{I}}$}
				\STATE{\textbf{return} $\mathbf{r}_k^{\ell,b}$}
			\end{algorithmic}
		\end{spacing}
	\end{algorithm}

	\subsection{Long-term Transmit Design Based on S-CSIT}
	In this subsection, we sketch an algorithm to solve the long-term problem in \eqref{op: PDD L} based on the CSSCA algorithm framework \cite{SSCA,TSSO}. Similar to the CSSCA algorithm for the GMM in Section \ref{section: Design of MA-Enhanced MIMO System}, an iterative algorithm is presented and in each iteration, the long-term variables $\mathbf{t}$ and $\left\{\mathbf{Q}_i\right\}$ are updated by solving a convex objective/feasibility surrogate problem obtained by replacing the objective and left hand-side terms of constraints C1$''_k$, $\forall k$, and C2 with their respective convex surrogate functions.
	
	In particular, in iteration $\ell$, the recursive concave surrogate function for the average achievable rate of UT $k$ is defined as $\hat f_{k}^{\ell}\left(\mathbf{t},\left\{\mathbf{Q}_i\right\}\right)$, which is constructed similarly to \eqref{eq: surrogate functions rate}. One difference is that, 
	we obtain the approximate gradients and the function values at $\mathbf{t}^{\ell}, \left\{\mathbf{Q}_i^{\ell}\right\}$ as shown in \eqref{eq: PDD gradients} at the top of  the next page according to the chain rule,
	where $\mathbf{r}_k^{\ell,b}$ is the stationary solution obtained from short-term problem \eqref{op: PDD S} for $\mathbf{H}_k^{\ell,b}$ and $\mathbf{U}\times_3\mathbf{V}$ is the tensor times matrix contraction along with the third mode of $\mathbf{U}$ to produce $\mathbf{V}$, i.e., $\left[\mathbf{U}\times_3\mathbf{V}\right]_{i,j,k}=\sum_l \left[\mathbf{U}\right]_{i,j,l}\left[\mathbf{V}\right]_{k,l}$.
	Then, the optimal solution to the corresponding approximate objective/feasibility problem in iteration $\ell$, $\bar{\mathbf{t}}^{\ell}$ and $\left\{\bar{\mathbf{Q}}_i^{\ell}\right\}$, is obtained by solving problem \eqref{op: surrogate 1} with $f_{k}^{\ell}\left(\mathbf{t},\left\{\mathbf{Q}_i\right\}\right)$ being replaced by $\hat f_{k}^{\ell}\left(\mathbf{t},\left\{\mathbf{Q}_i\right\}\right)$ (if \eqref{op: surrogate 1} is feasible). If problem \eqref{op: surrogate 1} is not feasible, feasibility problem \eqref{op: surrogate 2} with $f_{k}^{\ell}\left(\mathbf{t},\left\{\mathbf{Q}_i\right\}\right)$ being replaced by $\hat f_{k}^{\ell}\left(\mathbf{t},\left\{\mathbf{Q}_i\right\}\right)$ is solved instead. After obtaining $\bar{\mathbf{t}}^{\ell}$ and $\left\{\bar{\mathbf{Q}}_i^{\ell}\right\}$, the optimization variables are updated according to \eqref{eq: update t} and \eqref{eq: update Q}, respectively.
	
	The gradient information $\nabla_{\mathbf{t}}\tilde{\mathbf{r}}_k^{\hat{\ell}}$ and $\nabla_{\mathbf{Q}_i}\tilde{\mathbf{r}}_k^{\hat{\ell}}$ in \eqref{eq: PDD gradients} can be extracted via the deep unrolling technique \cite{TSSO}. In particular, these gradients can be calculated according to \eqref{eq: rk gradient wrt t Q} at the top of the next page. Here, (a) holds due to projection, i.e., a small enough $\mathbf{t}$ or $\left\{\mathbf{Q}_i\right\}$ cannot change $\left[\left[\tilde{\mathbf{r}}_k^{\hat{\ell}-1}+\tau^{\hat{\ell}}\nabla_{\mathbf{r}_k} R_k\left(\mathbf{t},\tilde{\mathbf{r}}_k^{\hat{\ell}-1},\left\{\mathbf{Q}_i\right\},\mathbf{H}_k\right) \right]_{\mathcal{C}_{r,k}}\right]_j$ when  $\left[\tilde{\mathbf{r}}_k^{\hat{\ell}-1}+\tau^{\hat{\ell}}\nabla_{\mathbf{r}_k} R_k\left(\mathbf{t},\tilde{\mathbf{r}}_k^{\hat{\ell}-1},\left\{\mathbf{Q}_i\right\},\mathbf{H}_k\right)\right]_j$ is outside of the $j$-th dimensional domain of $\mathcal{C}_{r,k}$. (b) holds as a result of the chain rule.
	\begin{figure*}[t]
				\begin{equation}
			\label{eq: PDD gradients}
			\begin{aligned}
				\hat{\mathbf{\Delta}}_{\mathbf{t},k}^{\ell} &\triangleq \frac{1}{B} 	\sum\nolimits_{b=1}^B 	\nabla_{\mathbf{t}} R_k\left(\mathbf{t}^{\ell}, \mathbf{r}_k^{\ell,b},\left\{\mathbf{Q}_i^{\ell}\right\},\mathbf{H}_k^{\ell,b}\right)+\nabla_{\mathbf{t}}\tilde{\mathbf{r}}_k^{\hat{I}}\left(\mathbf{t}^{\ell},\left\{\mathbf{Q}_i^{\ell}\right\},\mathbf{H}_k^{\ell,b}\right)\nabla_{\mathbf{r}_k} R_k\left(\mathbf{t}^{\ell}, \mathbf{r}_k^{\ell,b},\left\{\mathbf{Q}_i^{\ell}\right\},\mathbf{H}_k^{\ell,b}\right), \\
				\hat{\mathbf{\Delta}}_{\mathbf{Q}_i,k}^{\ell} &\triangleq 	\frac{1}{B} 	\sum\nolimits_{b=1}^B \nabla_{\mathbf{Q}_i} R_k\left(\mathbf{t}^{\ell}, \mathbf{r}_k^{\ell,b},\left\{\mathbf{Q}_i^{\ell}\right\},\mathbf{H}_k^{\ell,b}\right)+\nabla_{\mathbf{Q}_i}\tilde{\mathbf{r}}_k^{\hat{I}}\left(\mathbf{t}^{\ell},\left\{\mathbf{Q}_i^{\ell}\right\},\mathbf{H}_k^{\ell,b}\right) \times_3 \nabla_{\mathbf{r}_k}^T R_k\left(\mathbf{t}^{\ell}, \mathbf{r}_k^{\ell,b},\left\{\mathbf{Q}_i^{\ell}\right\},\mathbf{H}_k^{\ell,b}\right), \\
				\hat{R}_k^{\ell} &\triangleq \frac{1}{B} \sum\nolimits_{b=1}^B 		R_k\left(\mathbf{t}^{\ell}, \mathbf{r}_k^{\ell,b},\left\{\mathbf{Q}_i^{\ell}\right\},\mathbf{H}_k^{\ell,b}\right).
			\end{aligned}
		\end{equation}
		\begin{equation}
			\label{eq: rk gradient wrt t Q}
			\begin{aligned}
				\nabla_{\mathbf{t}}\tilde{\mathbf{r}}_k^{\hat{\ell}} &= \nabla_{\mathbf{t}}\left( \left[\tilde{\mathbf{r}}_k^{\hat{\ell}-1}+\tau^{\hat{\ell}}\nabla_{\mathbf{r}_k} R_k\left(\mathbf{t},\tilde{\mathbf{r}}_k^{\hat{\ell}-1},\left\{\mathbf{Q}_i\right\},\mathbf{H}_k\right) \right]_{\mathcal{C}_{r,k}} \right)
				\overset{\text{(a)}}{=} \left( \nabla_{\mathbf{t}} \tilde{\mathbf{r}}_k^{\hat{\ell}-1}+\tau^{\hat{\ell}}\nabla_{\mathbf{t}} \nabla_{\mathbf{r}_k} R_k\left(\mathbf{t},\tilde{\mathbf{r}}_k^{\hat{\ell}-1},\left\{\mathbf{Q}_i\right\},\mathbf{H}_k\right) \right) \mathbf{B}_k^{\hat{\ell}} \\
				&\overset{\text{(b)}}{=} \underbrace{\left( \nabla_{\mathbf{t}}\tilde{\mathbf{r}}_k^{\hat{\ell}-1} + \tau^{\hat{\ell}} \nabla_{\mathbf{t}} \tilde{\mathbf{r}}_k^{\hat{\ell}-1} \nabla_{\mathbf{r}_k}^2R_k\left(\mathbf{t},\tilde{\mathbf{r}}_k^{\hat{\ell}-1},\left\{\mathbf{Q}_i\right\},\mathbf{H}_k\right)+ \tau^{\hat{\ell}} 	\frac{\partial}{\partial\mathbf{t}}\nabla_{\mathbf{r}_k}R_k\left(\mathbf{t},\tilde{\mathbf{r}}_k^{\hat{\ell}-1},\left\{\mathbf{Q}_i\right\},\mathbf{H}_k\right) \right)}_{\triangleq\mathbf{C}_{t,k}^{\hat{\ell}}} \mathbf{B}_k^{\hat{\ell}} \in \mathbb{R}^{N\times M}, \\
				\nabla_{\mathbf{Q}_i}\tilde{\mathbf{r}}_k^{\hat{\ell}} &= \nabla_{\mathbf{Q}_i}\left( \left[\tilde{\mathbf{r}}_k^{\hat{\ell}-1}+\tau^{\hat{\ell}}\nabla_{\mathbf{r}_k} R_k\left(\mathbf{t},\tilde{\mathbf{r}}_k^{\hat{\ell}-1},\left\{\mathbf{Q}_i\right\},\mathbf{H}_k\right) \right]_{\mathcal{C}_{r,k}} \right)
				\overset{\text{(a)}}{=} \left( \nabla_{\mathbf{Q}_i} \tilde{\mathbf{r}}_k^{\hat{\ell}-1}+\tau^{\hat{\ell}}\nabla_{\mathbf{Q}_i} \nabla_{\mathbf{r}_k} R_k\left(\mathbf{t},\tilde{\mathbf{r}}_k^{\hat{\ell}-1},\left\{\mathbf{Q}_i\right\},\mathbf{H}_k\right) \right) \times_3 \mathbf{B}_k^{\hat{\ell}} \\
				&\overset{\text{(b)}}{=} \underbrace{\left( 	\nabla_{\mathbf{Q}_i}\tilde{\mathbf{r}}_k^{\hat{\ell}-1} + \tau^{\hat{\ell}} \nabla_{\mathbf{Q}_i} \tilde{\mathbf{r}}_k^{\hat{\ell}-1} \times_3 \nabla_{\mathbf{r}_k}^2R_k\left(\mathbf{t},\tilde{\mathbf{r}}_k^{\hat{\ell}-1},\left\{\mathbf{Q}_i\right\},\mathbf{H}_k\right)+ \tau^{\hat{\ell}} \frac{\partial}{\partial\mathbf{Q}_i} \nabla_{\mathbf{r}_k} R_k\left(\mathbf{t},\tilde{\mathbf{r}}_k^{\hat{\ell}-1},\left\{\mathbf{Q}_i\right\},\mathbf{H}_k\right) \right)}_{\triangleq\mathbf{C}_{Q,k,i}^{\hat{\ell}}} \times_3 \mathbf{B}_k^{\hat{\ell}} \in \mathbb{C}^{N\times N\times M}.
			\end{aligned}
		\end{equation}
		\hrule
	\end{figure*}
	Moreover, we  initialize $\tilde{\mathbf{r}}_k$ with a constant value. Thus, $\tilde{\mathbf{r}}_k^0$ is independent of $\mathbf{t}$ and $\mathbf{Q}_i$, i.e., $\nabla_{\mathbf{t}}\tilde{\mathbf{r}}_k^0=\mathbf{0}$, $\nabla_{\mathbf{Q}_i}\tilde{\mathbf{r}}_k^0=\mathbf{0}$. Combining \eqref{eq: short-term KKT} and \eqref{eq: rk gradient wrt t Q}, the terms in \eqref{eq: PDD gradients} converge to $0$ when $\hat{I} \rightarrow \infty$ as follows 
		\begin{equation}
			\label{eq: PDD gradients error}
			\begin{aligned}
				&\left\| 	\nabla_{\mathbf{t}}\tilde{\mathbf{r}}_k^{\hat{I}}\nabla_{\mathbf{r}_k} R_k\left(\mathbf{t}, \tilde{\mathbf{r}}_k^{\hat{I}},\left\{\mathbf{Q}_i\right\},\mathbf{H}_k\right) \right\| \\
				=& \left\| \mathbf{C}_{t,k}^{\hat{I}} \mathbf{B}_k^{\hat{I}}\nabla_{\mathbf{r}_k} R_k\left(\mathbf{t}, \tilde{\mathbf{r}}_k^{\hat{I}},\left\{\mathbf{Q}_i\right\},\mathbf{H}_k\right) \right\| \\
				\overset{(a)}{\leq}& \left\| \mathbf{C}_{t,k}^{\hat{I}} \right\| \left\| \mathbf{B}_k^{\hat{I}}\nabla_{\mathbf{r}_k} R_k\left(\mathbf{t}, \tilde{\mathbf{r}}_k^{\hat{I}},\left\{\mathbf{Q}_i\right\},\mathbf{H}_k\right) \right\| \\
				=& \mathcal{O}\left(e^{\hat{I}}\left(\mathbf{t},\left\{\mathbf{Q}_i\right\}\right)\right), \\
				&\left\| 	\nabla_{\mathbf{Q}_i}\tilde{\mathbf{r}}_k^{\hat{I}}\times_3\nabla_{\mathbf{r}_k}^T R_k\left(\mathbf{t}, \tilde{\mathbf{r}}_k^{\hat{I}},\left\{\mathbf{Q}_i\right\},\mathbf{H}_k\right) \right\| \\
				=& \left\| \mathbf{C}_{Q,k,i}^{\hat{I}} \times_3 \mathbf{B}_k^{\hat{I}} \times_3 \nabla_{\mathbf{r}_k}^T R_k\left(\mathbf{t}, \tilde{\mathbf{r}}_k^{\hat{I}},\left\{\mathbf{Q}_i\right\},\mathbf{H}_k\right) \right\| \\
				=& \left\| \mathbf{C}_{Q,k,i}^{\hat{I}} \times_3 \left(\mathbf{B}_k^{\hat{I}} \nabla_{\mathbf{r}_k} R_k\left(\mathbf{t}, \tilde{\mathbf{r}}_k^{\hat{I}},\left\{\mathbf{Q}_i\right\},\mathbf{H}_k\right)\right)^T \right\| \\
				\overset{(a)}{\leq}& \left\| \mathbf{C}_{Q,k,i}^{\hat{I}} \right\| \left\| \mathbf{B}_k^{\hat{I}} \nabla_{\mathbf{r}_k} R_k\left(\mathbf{t}, \tilde{\mathbf{r}}_k^{\hat{I}},\left\{\mathbf{Q}_i\right\},\mathbf{H}_k\right) \right\| \\
				=& \mathcal{O}\left(e^{\hat{I}}\left(\mathbf{t},\left\{\mathbf{Q}_i\right\}\right)\right),
			\end{aligned}
		\end{equation}
		where (a) holds since the Frobenius norm is submultiplicative.
		Based on the proof of \cite[Theorems 2 and 3]{TSSO}, we can omit these terms in practice without introducing any adverse effects on the convergence of the PDD-SSCA algorithm.
	
	\begin{algorithm}[t]
		\caption{PDD-SSCA Algorithm for Solving Problem \eqref{op: original 2}}
		\label{Al: PDD SSCA t Q}
		\begin{spacing}{1}
			\textbf{Input:} $N$, $M$, $K$, $\sigma^2$, $\lambda$, $\mathcal{C}_{t}$, $\{\mathcal{C}_{r,k}\}$; $\left\{\tau_{t,k}\right\}$, $\left\{\tau_{Q,k,i}\right\}$, $\left\{\tau_{h_{i,j}}\right\}$, $\hat{I}$; S-CSIT \\
			\textbf{Output:} $\mathbf{t}^*$, $\left\{\mathbf{Q}_i^*\right\}$
			\begin{algorithmic}[1]
				\STATE {Initialize $\mathbf{t}^1$ and $\left\{\mathbf{Q}_i^1\right\}$ satisfying the position and power constraints, respectively}
				\FOR{$\ell = 1 \text{ to } I$}
				\STATE {Obtain a mini-batch $\left\{ \mathbf{H}_k^{\ell,b}, b=1,\cdots,B \right\}$ and then compute $\left\{\mathbf{r}_k^{\ell,b}\right\}$ by Algorithm \ref{Al: Rx r GP}}
				\STATE {Construct the surrogate functions $\hat f_k^{\ell}\left(\mathbf{t},\left\{\mathbf{Q}_i\right\}\right)$, $\forall k$ with approximate gradients and the function values given by \eqref{eq: PDD gradients} and replace $f_{k}^{\ell}\left(\mathbf{t},\left\{\mathbf{Q}_i\right\}\right)$, $\forall k$ in problems \eqref{op: surrogate 1} and \eqref{op: surrogate 2} with $\hat f_{k}^{\ell}\left(\mathbf{t},\left\{\mathbf{Q}_i\right\}\right)$, $\forall k$}
				\IF{problem \eqref{op: surrogate 1}  is feasible}
				\STATE {Obtain $\bar{\mathbf{t}}^{\ell}$ and $\left\{\bar{\mathbf{Q}}_i^{\ell}\right\}$ by solving \eqref{op: surrogate 1}}
				\ELSE
				\STATE {Obtain $\bar{\mathbf{t}}^{\ell}$ and $\left\{\bar{\mathbf{Q}}_i^{\ell}\right\}$ by solving \eqref{op: surrogate 2}}
				\ENDIF
				\STATE{Update $\mathbf{t}^{\ell+1}$ and $\left\{\mathbf{Q}_i^{\ell+1}\right\}$ according to \eqref{eq: update variables}}
				\ENDFOR
				\STATE {Set $\mathbf{t}^*=\mathbf{t}^{I+1}$, $\left\{\mathbf{Q}_i^*=\mathbf{Q}^{I+1}_i\right\}$}
				\STATE{\textbf{return} $\mathbf{t}^*$, $\left\{\mathbf{Q}_i^*\right\}$}
			\end{algorithmic}
		\end{spacing}
	\end{algorithm}
	
	The overall PDD-SSCA algorithm framework  for solving problem \eqref{op: original 2} is summarized in Algorithm \ref{Al: PDD SSCA t Q}. In Step 1, the transmit APV $\mathbf{t}^1$ is initialized to satisfy the position and power constraints, respectively.
	In Steps 2-11, the transmit APV and covariance matrices are jointly optimized using the CSSCA method. The convergence of Algorithm \ref{Al: PDD SSCA t Q} is guaranteed by  \cite[Theorems 2 and 3]{TSSO}. In particular, every limit point of the sequence $\{\mathbf{t}^{\ell},\left\{\mathbf{Q}_i^{\ell}\right\}\}_{\ell=1}^\infty$ generated by Algorithm \ref{Al: PDD SSCA t Q} almost surely satisfies the KKT conditions of problem \eqref{op: original 2} up to an error of $ \mathcal{O}\left(e^{\hat{I}}\left(\mathbf{t},\left\{\mathbf{Q}_i\right\}\right)\right)$, where $\lim _{\hat{I} \rightarrow \infty} e^{\hat{I}}(\mathbf{t},\left\{\mathbf{Q}_i\right\})=0$. On the other hand, the complexity of Algorithm \ref{Al: PDD SSCA t Q} is analyzed as follows. Firstly, obtaining a series of stationary point to problem \eqref{op: PDD S}, $\left\{\mathbf{r}_k^{\ell,b}\right\}$, entails complexity $\mathcal{O}\left(BK\hat{I}M^3\right)$. Secondly, the complexity of calculating $\nabla_{\mathbf{t}} R_k(\mathbf{t}^{\ell}, \mathbf{r}_k^{\ell,b},\left\{\mathbf{Q}_i^{\ell}\right\},\mathbf{H}_k^{\ell,b})$ and $\nabla_{\mathbf{Q}_i} R_k(\mathbf{t}^{\ell}, \mathbf{r}_k^{\ell,b},\left\{\mathbf{Q}_i^{\ell}\right\},\mathbf{H}_k^{\ell,b})$, $i=1,2,\cdots,K$, $\forall k$, is $\mathcal{O}\left(K M^{3}\right)$, respectively. Finally, the complexity of obtaining the solution for each objective/feasibility surrogate problem by the interior-point method with accuracy $\epsilon$ is $\mathcal{O}\left(\log(1/\epsilon)N^{7}\right)$ \cite{CO}. Thus, the complexity of Algorithm \ref{Al: PDD SSCA t Q} is
	$\mathcal{O}\left(I\left(BK\hat{I}M^3+\log(1/\epsilon)N^{7}\right)\right)$.
	
	\section{Numerical Results}
	
	In this section, we numerically evaluate the performance of the proposed multiuser MA-enhanced MIMO system. We first introduce the simulation setup and benchmark schemes. Subsequently, we present numerical results to verify the efficacy of the proposed algorithms and the two movement modes.
	
	\subsection{Simulation Setup and Benchmark Schemes}
	
	In our simulations, the UTs are randomly distributed around the BS with their distances w.r.t. the latter, $d_k$, following a uniform distribution between $20$ and $100$ meters (m). For UT $k$, the average channel gain is set as $g_k = c_0 d_k^{-\alpha_0}$, where $c_0$ denotes the expected value of the path loss at the reference distance of $1$ m, and $\alpha_0$ represents the path loss exponent.
	We assume that there are $L$ paths for each UT, $L_t=L_r=L$ and $\left[\mathbf{\Sigma}_{k}\right]_{l,l} \sim \mathcal{CN}\left(0,\frac{g_k}{L}\right)$, $1 \leq l \leq L$. Moreover, the $L$ pairs of elevation/azimuth AoDs and AoAs for UT $k$ are i.i.d. random variables following distributions $f\left(\theta_{t,k}^l,\phi_{t,k}^l\right)=\frac{1}{2\pi} \sin \phi_{t,k}^l$, $\theta_{t,k}^l\in\left[0,\pi\right]$, $\phi_{t,k}^l\in\left[0,\pi\right]$ and $f\left(\theta_{r,k}^l,\phi_{r,k}^l\right)=\frac{1}{2\pi} \sin \phi_{r,k}^l$, $\theta_{r,k}^l\in\left[0,\pi\right]$, $\phi_{r,k}^l\in\left[0,\pi\right]$, $1\leq l \leq L$, respectively. 
	Each point in the simulation figures was averaged over 1000 different random CSI realizations.
	The default settings for the simulation parameters are provided in Table \ref{Tb: simulation parameter} \cite{Mccf,SSCA,TSSO}. The sizes of the transmit regions for GMM and PMM are set as $\left(4X_t+3D\right)\times \left(2X_t+D\right)$, the size of the receive region for GMM is set as $\left(2X_r+D\right)\times X_r$ and that for PMM is consist of $2$ horizontal-arranged regions with size $X_r\times X_r$. 
	Besides, the transmit and receive MAs are initialized as UPAs with adjacent antennas spaced by $D+\frac{X_t}{2}$ and $D+\frac{X_r}{2}$, respectively. The transmit covariance matrices are initialized as diagonal matrices, where the main diagonal entries are identical and their summation is equal to $P$.
	\begin{table}[htb]
		\begin{center}
			\caption{Simulation Parameters.}
			\label{Tb: simulation parameter}
			\begin{tabular}{|c|c|c|c|}
				\hline Parameter & Value & Parameter & \,\, Value \,\, \\
				\hline$N$ & $8$ & $I$ & $100$ \\
				\hline$M$ & $2$ & $\tau_{t,k}$, $\tau_{h_{i,j}}$ & $-1$ \\
				\hline$K$ & $4$ & $\tau_{Q,k,i}$ & $-P^{-2}$ \\
				\hline$D$, $X_t$, $X_r$ & $0.5\lambda$ & $B$ & $10$ \\
				\hline $\lambda$ & $6\mathrm{~cm}$ & $\rho^{\ell}$ & $\left(\ell+1\right)^{-0.9}$ \\
				\hline $L$ & $10$ & $\gamma^{\ell}$ & $\left(\ell+1\right)^{-1}$ \\
				\hline $\sigma^2$ & $-80\mathrm{~dBm}$ & $s$ & $10$  \\
				
				\hline $c_0$ & $-40\mathrm{~dB}$ & $\tau$ & $0.5$ \\
				\hline $\alpha_0$ & $2.8$ & $\xi$ & $0.6$ \\
				\hline $P$ & $20 \,\, \mathrm{dBm}$ & $\epsilon$ & $10^{-6}$ \\
				\hline $R_{\mathrm{min}}$ & $1 \,\, \mathrm{bps/Hz}$ & $\hat{I}$ & 30 \\
				\hline 
			\end{tabular}
		\end{center}
	\end{table} 

	We compare the performance of the schemes proposed in Sections \ref{section: Design of MA-Enhanced MIMO System} (\textbf{Proposed GMM}) and \ref{section: Antenna Move Mode Design} (\textbf{Proposed PMM}) with the following schemes, where the CSSCA algorithm is tailored to tackle the resulting problems:
	\begin{itemize}
		\item [$\bullet$]\textbf{Decoupled-GMM}: The system setup in this scheme is set as the same as the proposed GMM scheme, except that Algorithms \ref{Al: Rx r GA} and \ref{Al: CSSCA t r Q} are decoupled, i.e., $\mathbf{r}_k^{\ell,b}$ in \eqref{eq: gradients} for Algorithm \ref{Al: CSSCA t r Q} is replaced by a constant that is set to the initial point of Algorithm \ref{Al: Rx r GA}.
		\item [$\bullet$]\textbf{S-CSIT-GMM}: The receive APVs are jointly optimized with the transmit APV and the transmit covariance matrices based on S-CSIT only. 
		\item [$\bullet$]\textbf{S-CSIT-UPA}: The BS and each UT are equipped with $4\times2$ and $2\times1$ UPAs, respectively, where adjacent antennas are spaced by $0.5\lambda$ unless specified otherwise. The transmit covariance matrices are optimized based on S-CSIT only. 
	\end{itemize}
	
	\subsection{Convergence Behavior of Proposed Algorithms}
	\begin{figure}
		\centering
		\subfigure[Achievable rate versus the number of iterations. The dashed line represents the convergence behavior of Algorithm \ref{Al: Rx r GA} when running for $30$ iterations.] {\includegraphics[width=0.44\textwidth,height=0.34\textwidth]{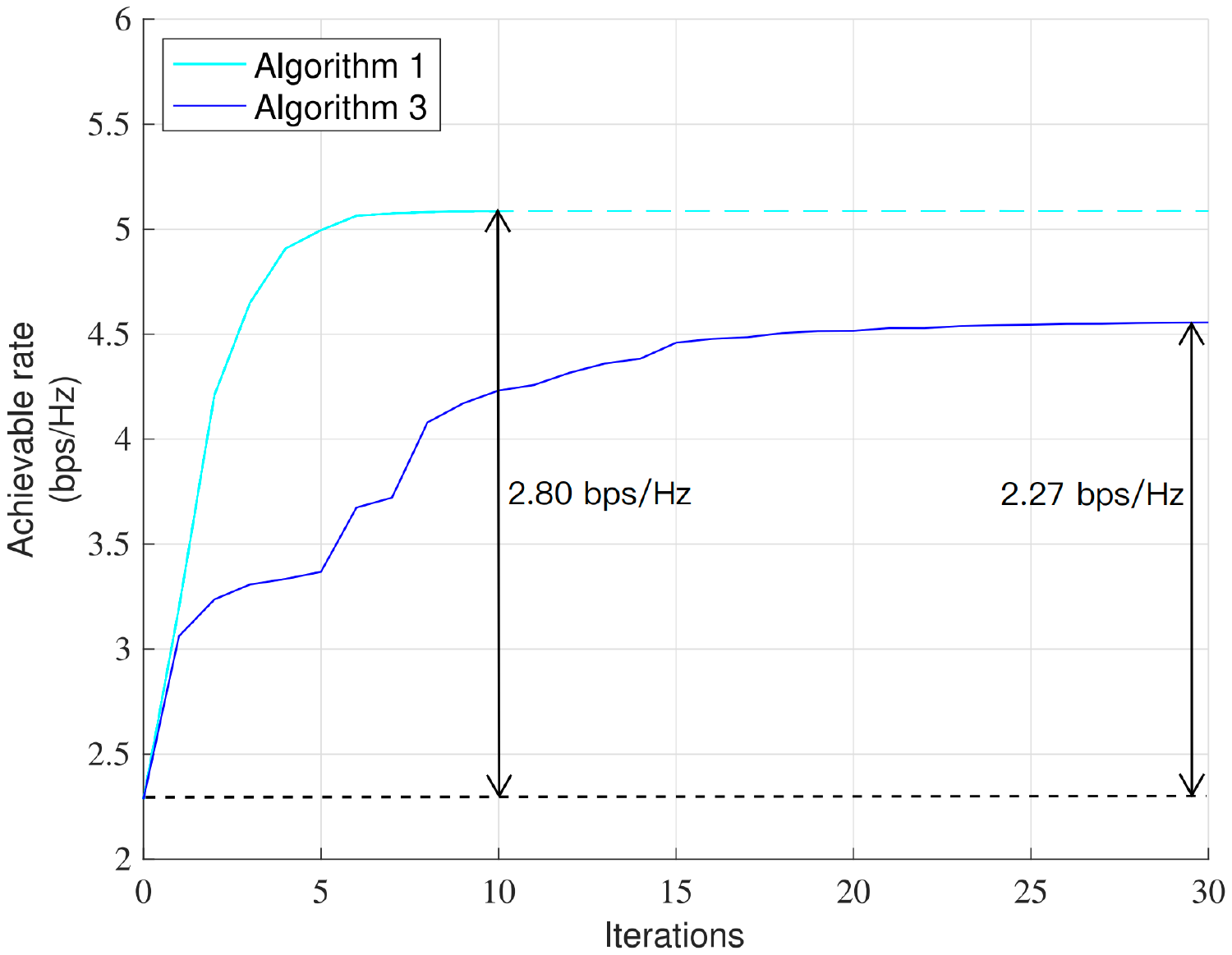}\label{fig: Iter}}
		\subfigure[Average achievable sum rate and minimum rate versus the number of iterations.]{\includegraphics[width=0.44\textwidth,height=0.34\textwidth]{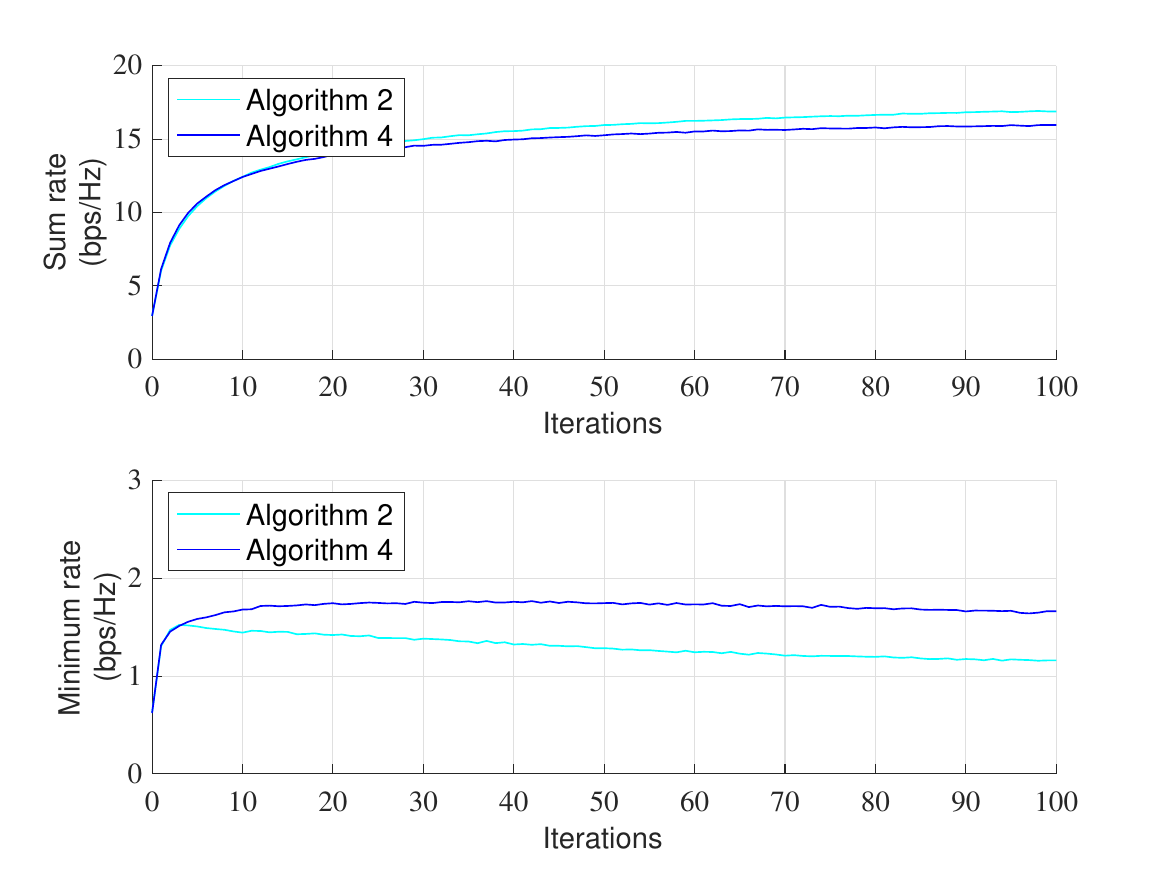}\label{fig: Iter_}}
		\caption{Convergence behavior of the proposed algorithms.}
	\end{figure}
	Fig.~\ref{fig: Iter} illustrates the convergence behavior of Algorithms \ref{Al: Rx r GA} and \ref{Al: Rx r GP}. It is observed that the achievable rates converge to values around $5.09$ and $4.56$ bps/Hz after $30$ iterations, which demonstrates the rapid convergence of the two algorithms. Note that Algorithm \ref{Al: Rx r GA} terminates after $10$ iterations when the incremental improvement in the objective function over two consecutive iterations falls below $\epsilon$. Additionally, Algorithms \ref{Al: Rx r GA} and \ref{Al: Rx r GP}  yield gains of $2.80$ and $2.27$ bps/Hz in the achievable rate, respectively. Fig.~\ref{fig: Iter_} shows the convergence behavior of Algorithms \ref{Al: CSSCA t r Q} and \ref{Al: PDD SSCA t Q} proposed, respectively. The average achievable sum rates gradually increase and converge to values around $16.88$ and $15.96$ bps/Hz after $100$ iterations. Moreover, since the proposed algorithms tend to reallocate resources from UTs with poor channel conditions to those with superior channel conditions, the minimum rates for both proposed algorithms gradually diminish and stabilize around the required minimum rate $R_{\mathrm{min}}$.
	
	\subsection{Impact of Maximum Transmit Power}
	\begin{figure*}
		\centering
		\subfigure[Average achievable sum rate.]{\includegraphics[width=0.44\textwidth,height=0.34\textwidth]{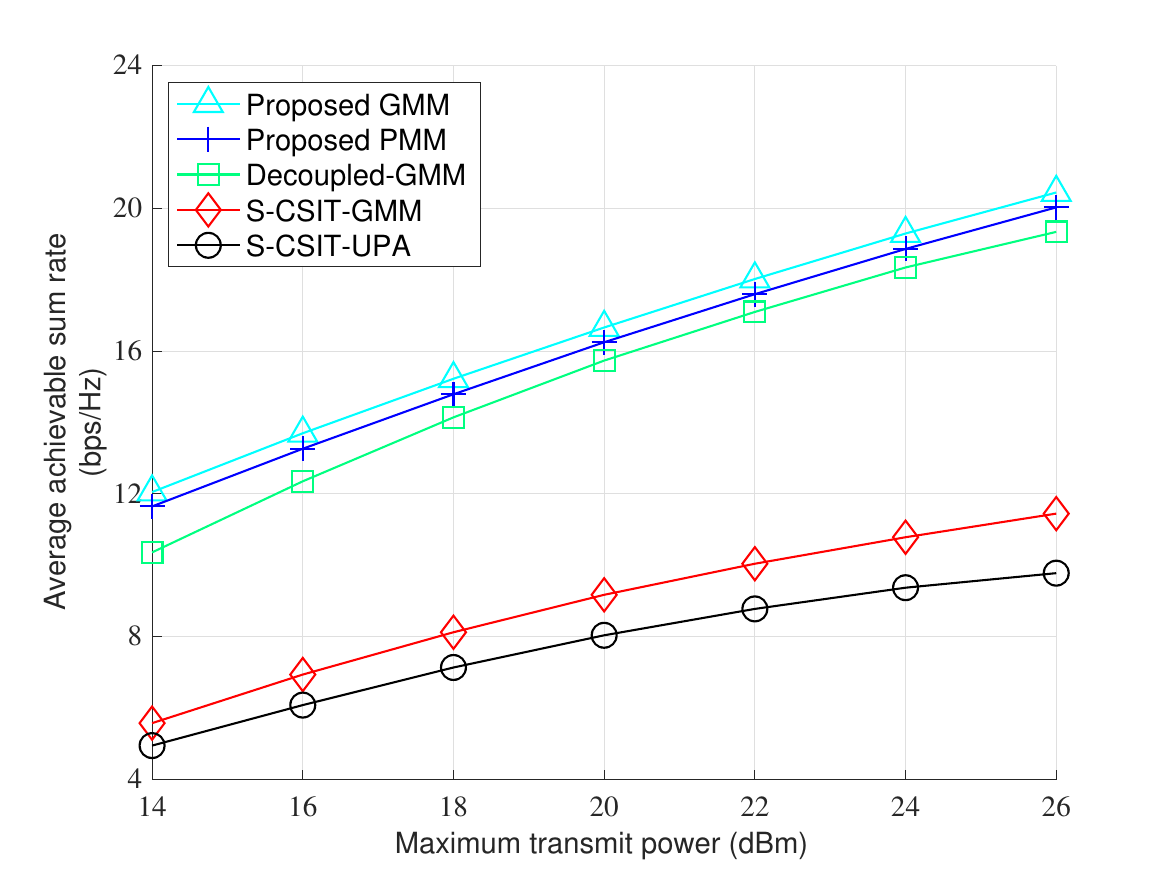}}
		\subfigure[Feasibility ratio.]{\includegraphics[width=0.44\textwidth,height=0.34\textwidth]{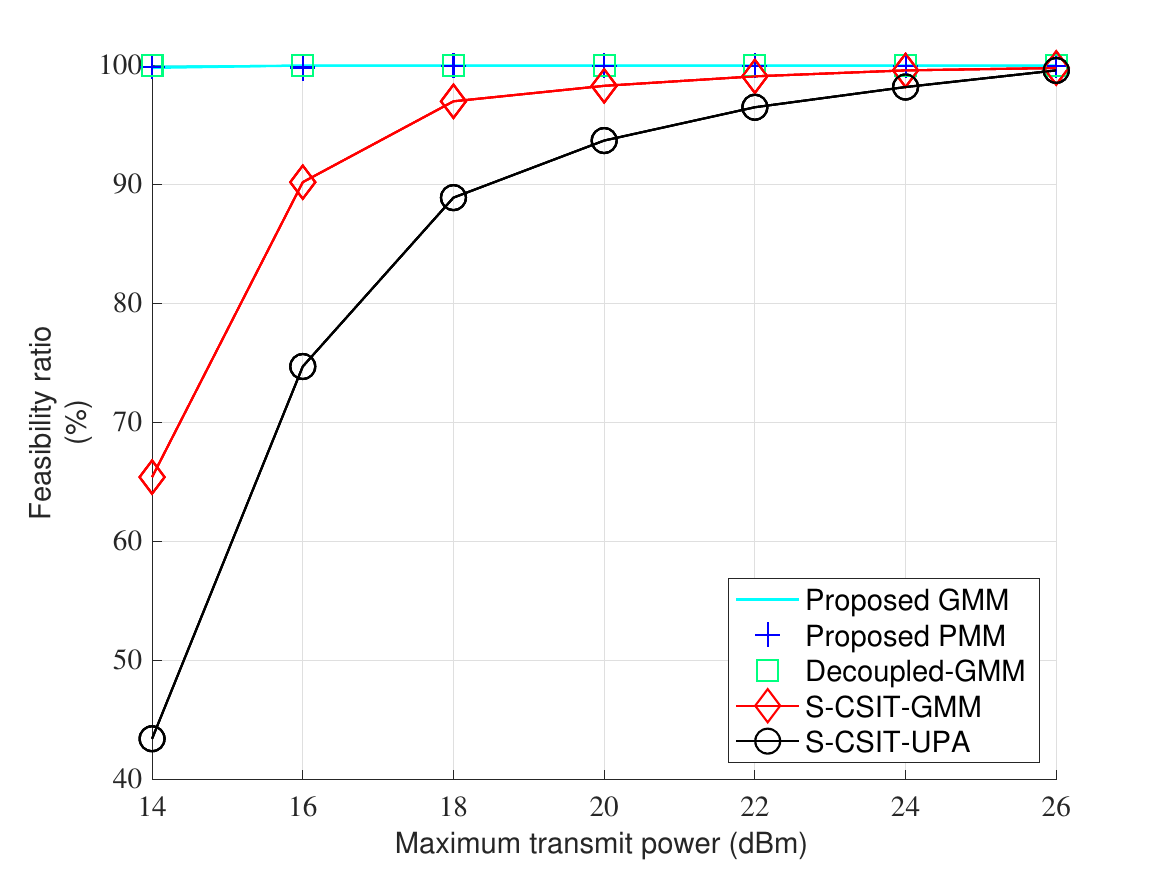}}
		\caption{Average achievable sum rate and feasibility ratio versus the maximum transmit power.}
		\label{fig: Simulation Power}
	\end{figure*}
	Fig.~\ref{fig: Simulation Power} presents the average achievable sum rate and the feasibility ratio (i.e., the proportion of solutions satisfying the minimum rate constraints) versus the maximum transmit power. As expected, for all considered systems, the average achievable sum rate is monotonically increasing in the maximum transmit power. Moreover, since the MAs can dynamically adapt to the channel conditions, the MA-enhanced MIMO systems outperform the conventional fixed-position UPA system. Since the receive APVs are designed based on I-CSIR, the proposed GMM, the proposed PMM, and the Decoupled-GMM schemes yield a significant improvement in both the average achievable sum rate and the feasibility ratio compared to the S-CSIT-GMM and S-CSIT-UPA schemes, which rely solely on S-CSIT. Notably, the feasibility ratio approaches nearly $100\%$ even for low transmit powers, indicating a considerable enhancement over the latter two schemes. For the Decoupled-GMM scheme, the long-term problem is solved by a modified version of Algorithm \ref{Al: CSSCA t r Q}, which does not account for the short-term receive APV design. In contrast, the proposed GMM and PMM schemes utilize Algorithms \ref{Al: Rx r GA} and \ref{Al: Rx r GP}, respectively, which optimize the receive APVs based on I-CSIR samples when constructing the surrogate problems of Algorithms \ref{Al: CSSCA t r Q} and \ref{Al: PDD SSCA t Q} for addressing the long-term problem. Indeed, this prevents excessive allocation of resources to UTs with poor channel conditions to satisfy their minimum rate constraints, thus enhancing performance relative to the Decoupled-GMM scheme. Furthermore, the proposed GMM scheme achieves the highest performance across all considered power budgets.
	
	\begin{figure*}
		\centering
		\subfigure[Average energy efficiency.]{\includegraphics[width=0.44\textwidth,height=0.34\textwidth]{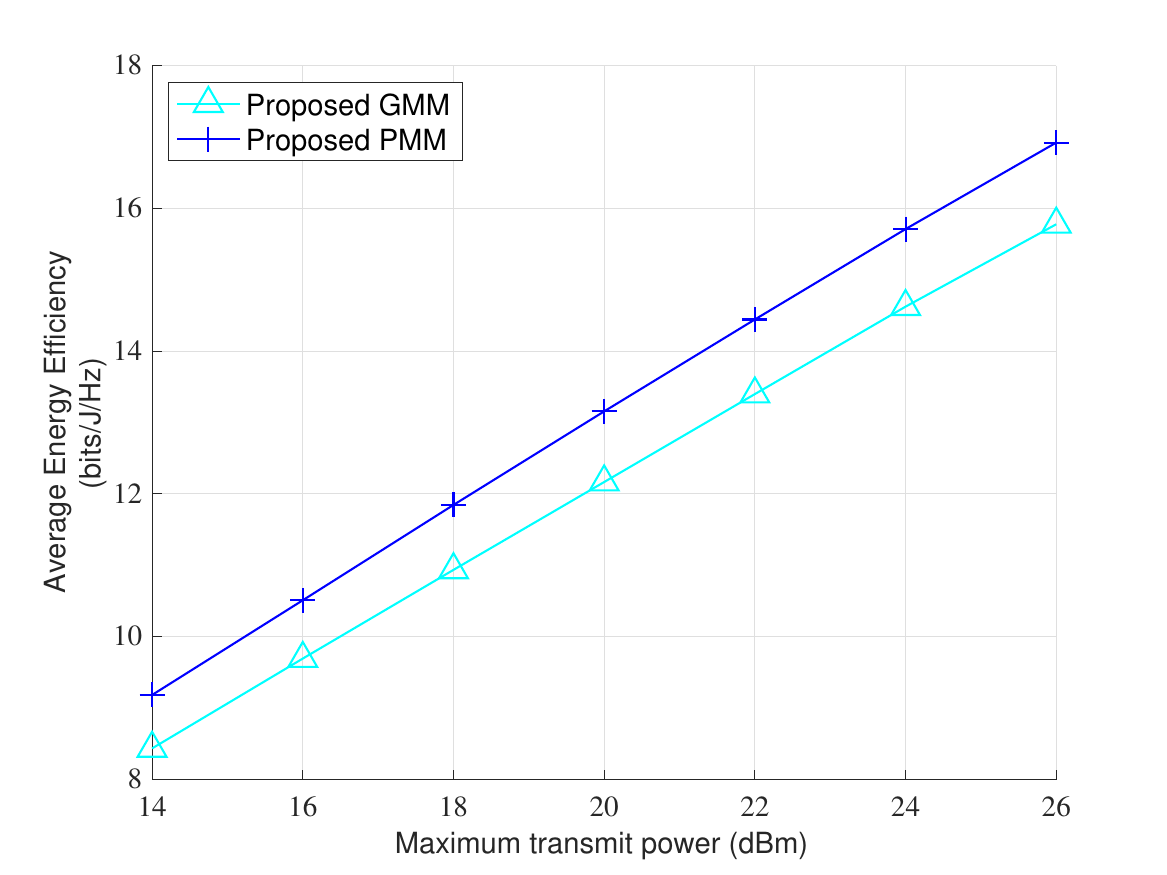}}
		\subfigure[Average repositioning delay.]{\includegraphics[width=0.44\textwidth,height=0.34\textwidth]{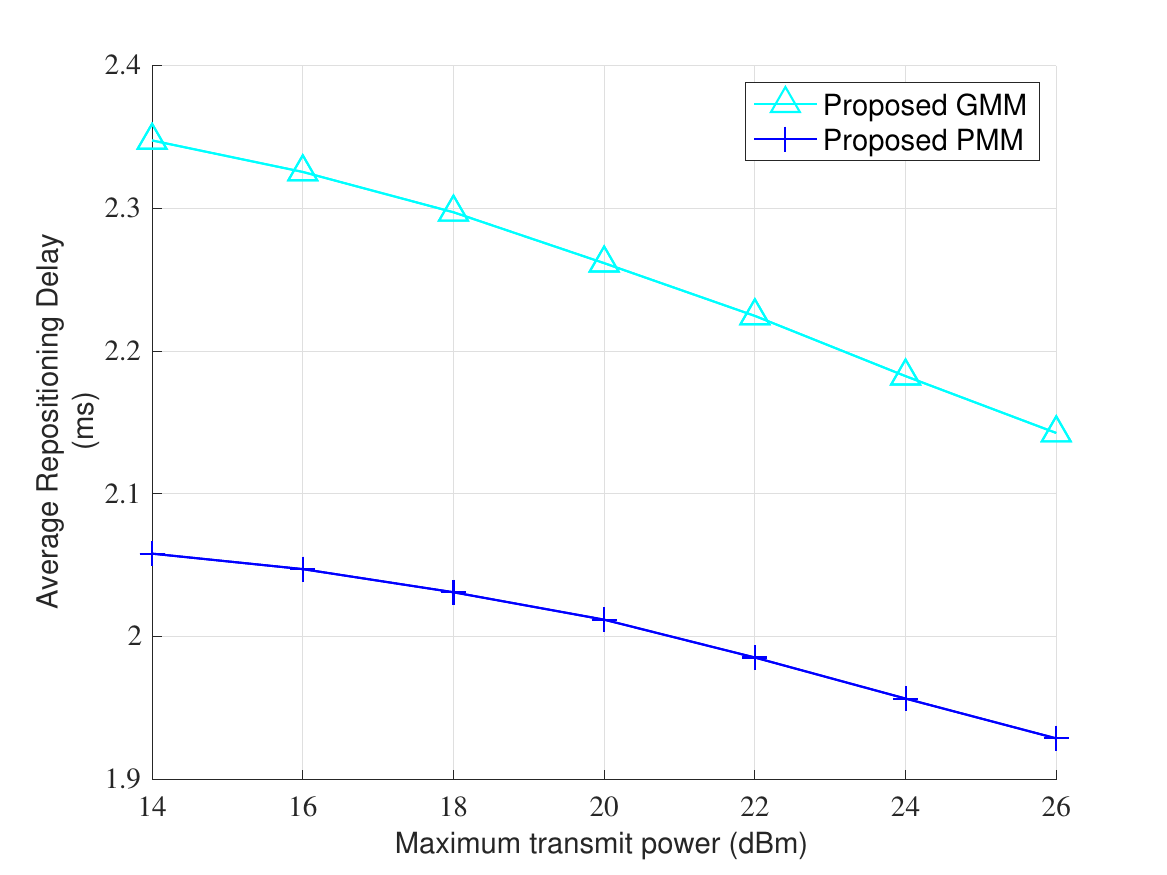}}
		\caption{Average energy efficiency and repositioning delay versus the maximum transmit power.}
		\label{fig: Simulation Comp GMMandPMM}
	\end{figure*}
	 Fig.~\ref{fig: Simulation Comp GMMandPMM} compares the effects of the receive antenna movement modes on the average energy efficiency and the repositioning delay. The energy efficiency is defined as the ratio of each UT's achievable rate to its repositioning power consumption. The power consumption is formulated as the product of a mobility-related coefficient (set to $1~\mathrm{J/m}$) and the total repositioning distances of each UT's receive MAs within one second, where the coherence time interval is set to $100~\mathrm{ms}$. Additionally, the repositioning delay is defined as the ratio of the maximum repositioning distance of each UT's receive MAs to the repositioning velocity (set to $10~\mathrm{m/s}$). As observed, the PMM scheme achieves significantly superior average energy efficiency and reduced repositioning delay compared to the GMM scheme, primarily attributed to its lower average and maximum repositioning distances, respectively.
	
	\subsection{Impact of Movement Region}
	\begin{figure}
		\centering
		\includegraphics[width=0.44\textwidth,height=0.34\textwidth]{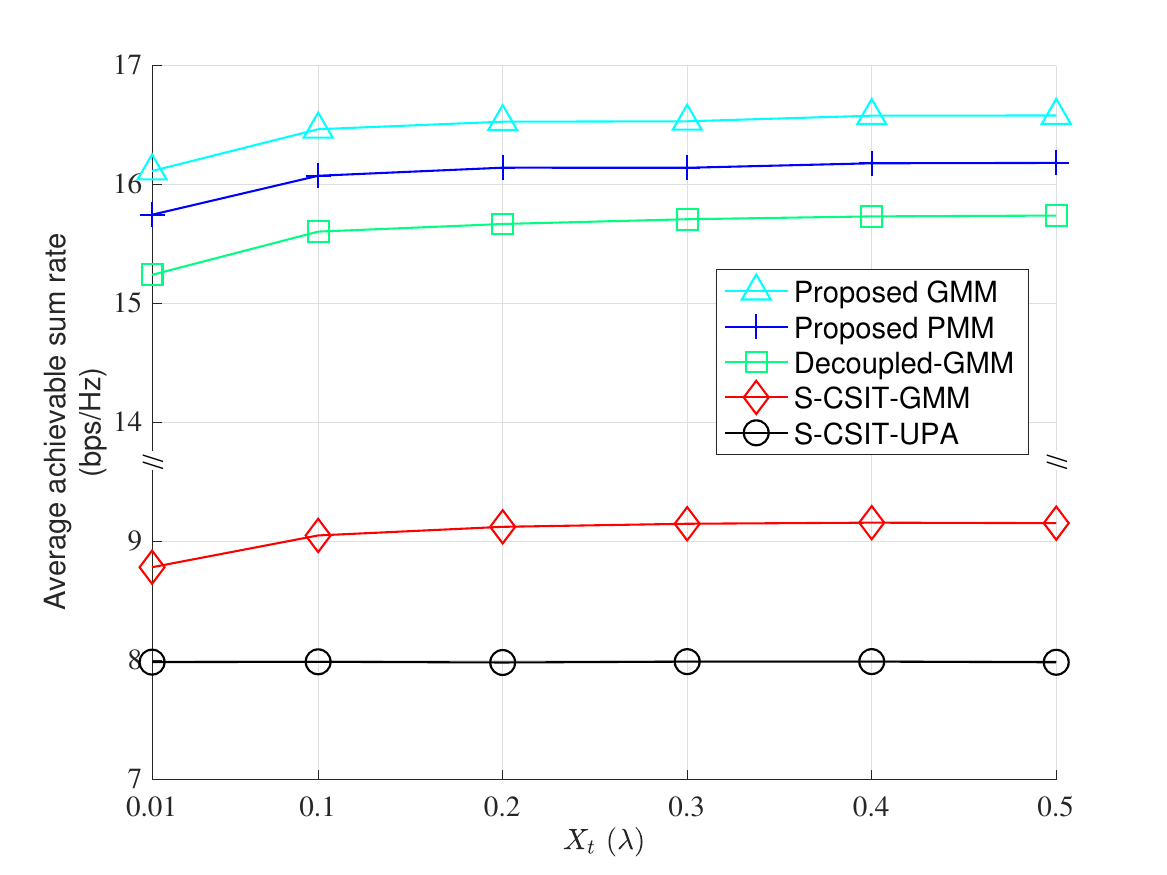}
		\caption{Average achievable sum rate versus the size of the transmit antenna movement region. For a specific $X_t$, the distance between adjacent transmit antennas for the S-CSIT-UPA benchmark scheme is set as $D+\frac{X_t}{2}$.}
		\label{fig: Simulation TSize}
	\end{figure}
	Fig. \ref{fig: Simulation TSize} depicts the relationship between the average achievable sum rate and the size of the transmit antenna movement region. Since the S-CSIT-UPA scheme lacks the capability to dynamically adapt the antenna positions based on the channel conditions, the average achievable sum rate does not depend on the inter-antenna distance. 
	In contrast, an increase in $X_t$ enables the MAs to more effectively exploit the long-term S-CSIT variations across a larger movement region. As a result, for the considered MA-enhanced MIMO systems, the average achievable sum rate increases with $X_t$ before it saturates.
	
	\begin{figure}
		\centering
		\includegraphics[width=0.44\textwidth,height=0.34\textwidth]{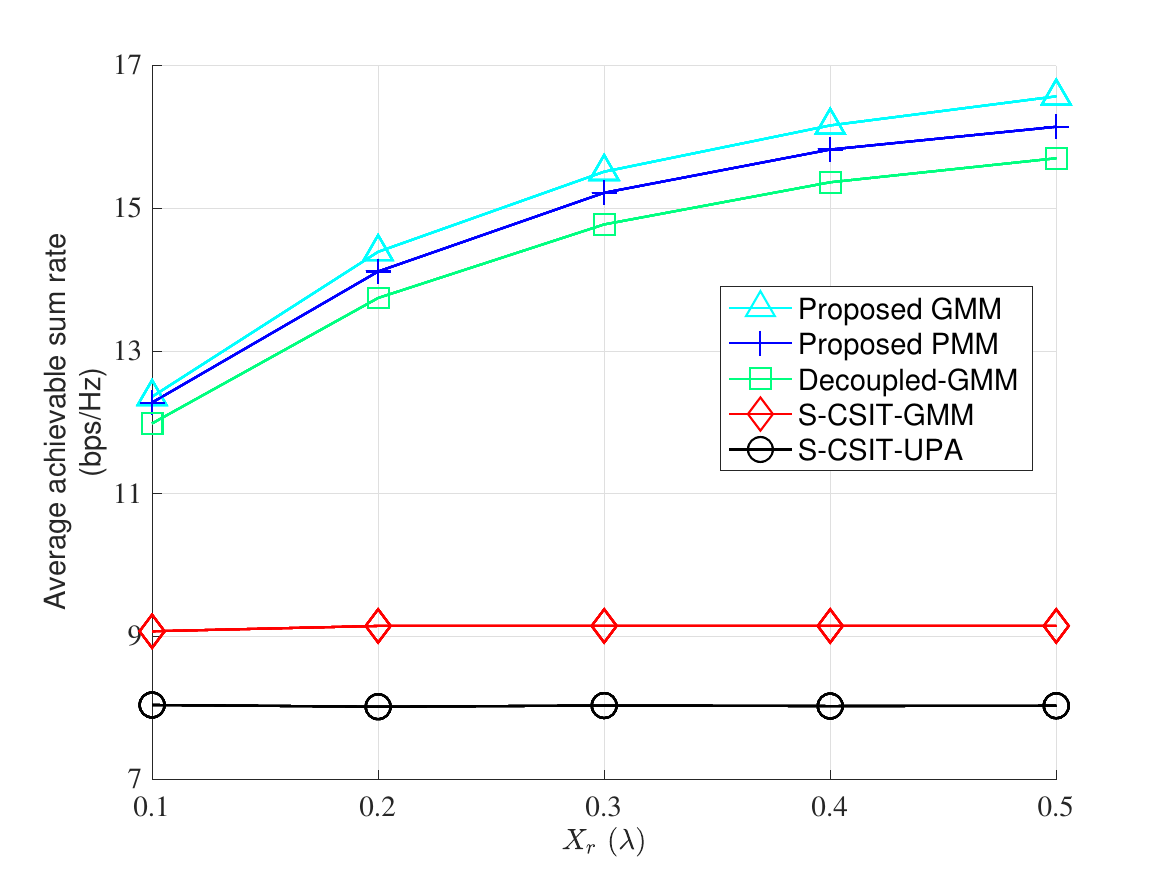}
		\caption{Average achievable sum rate versus the size of the receive antenna movement region. For a specific $X_r$, the distance between adjacent receive antennas for the S-CSIT-UPA benchmark scheme is set as $D+\frac{X_r}{2}$.}
		\label{fig: Simulation RSize}
	\end{figure}
	Fig. \ref{fig: Simulation RSize} depicts the relationship between the average achievable sum rate and the size of the receive antenna movement region. Since the exploited spatial DoFs increase, the average achievable sum rate for the considered MA-enhanced MIMO systems grows with $X_r$. In contrast, the average achievable sum rate for the S-CSIT-UPA benchmark scheme remains constant, as the receive APVs are unable to adapt to either S-CSIT or I-CSIR. As $X_r$ grows, the proposed GMM, the proposed PMM, and the Decoupled-GMM schemes yield a more substantial increase in the average achievable sum rate compared to the S-CSIT-GMM scheme. This improvement is attributed to the ability of the receive APVs for the former three schemes to adapt to short-term I-CSIR variations across the movement regions. Moreover, the performance gap between the proposed GMM and the proposed PMM schemes increases with $X_r$, since the size of the movement region for each receive MA for the PMM scheme expands at a slower rate, i.e., as $X_r\times X_r$, compared to that for the GMM scheme, i.e., as $\left(2X_r+D\right)\times X_r$.
	
	\begin{figure}
		\centering
		\includegraphics[width=0.44\textwidth,height=0.34\textwidth]{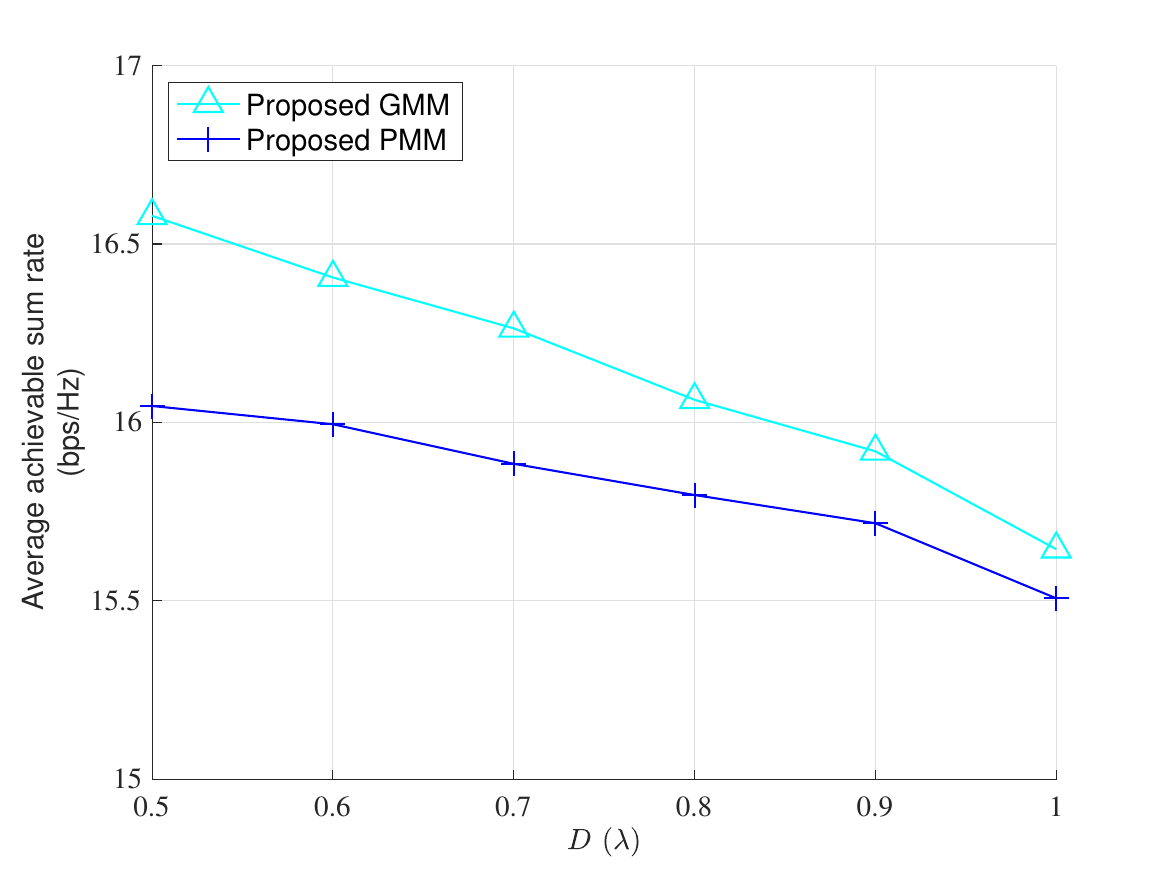}
		\caption{Average achievable sum rate versus the minimum distance between adjacent MAs.}
		\label{fig: Simulation D}
	\end{figure}
	Furthermore, Fig. \ref{fig: Simulation D} demonstrates that both the average achievable sum rate and the performance gap exhibit significant decreasing trends as the minimum distance between adjacent MAs increases. This performance degradation can be attributed to the progressively restricted antenna mobility.
	
	\subsection{Impact of Required Minimum Rate}
	\begin{figure}[!h]
		\centering
		\subfigure[Average achievable sum rate.]{\includegraphics[width=0.44\textwidth,height=0.34\textwidth]{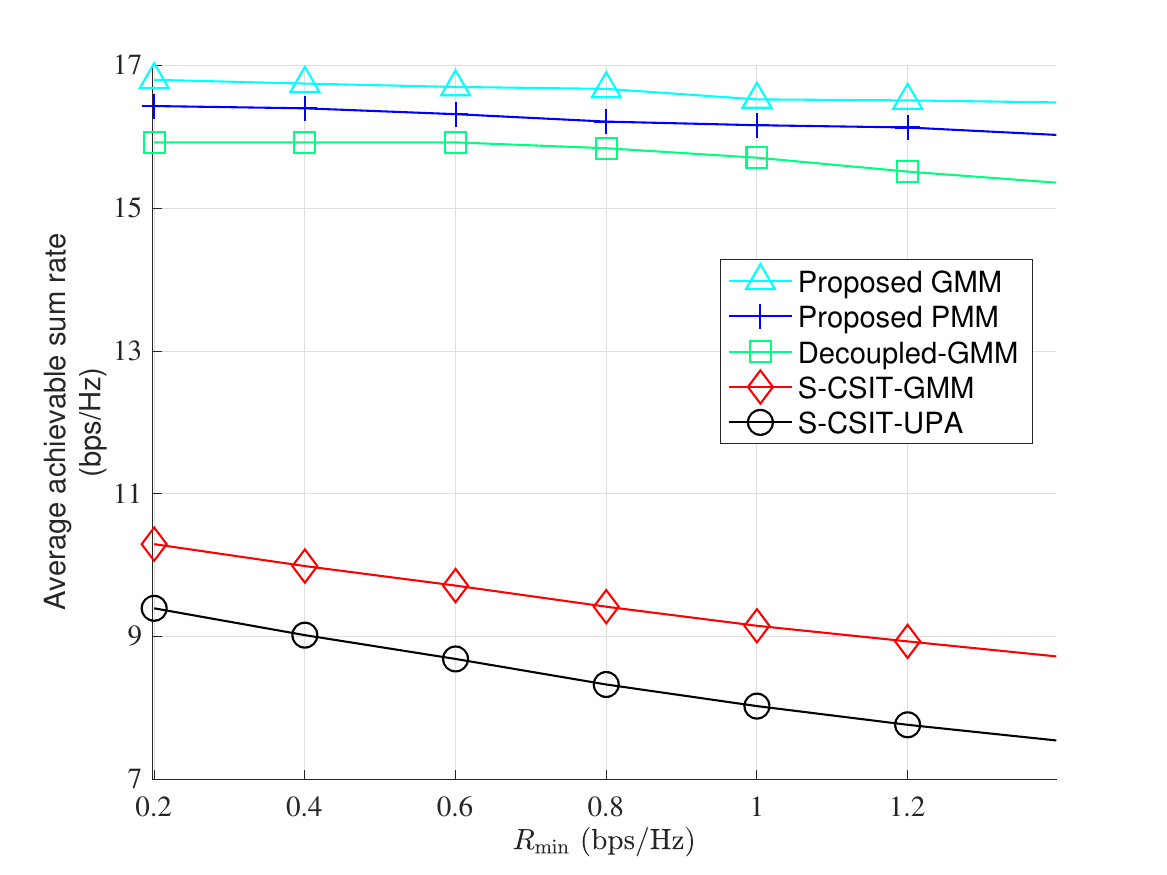}}
		\subfigure[Feasibility ratio.]{\includegraphics[width=0.44\textwidth,height=0.34\textwidth]{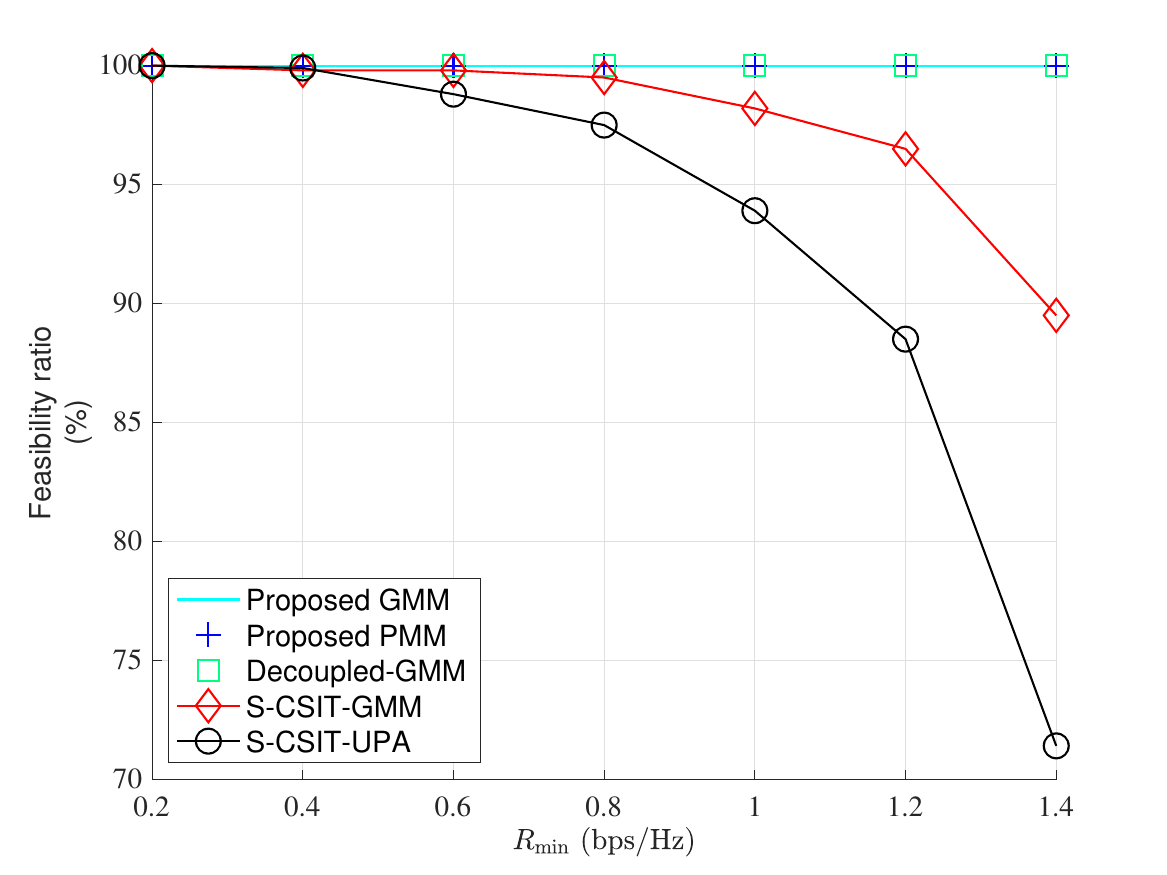}}
		\caption{Average achievable sum rate and feasibility ratio versus the required minimum rate.}
		\label{fig: Simulation Rmin}
	\end{figure}	
	Fig. \ref{fig: Simulation Rmin} illustrates the average achievable sum rate and the feasibility ratio versus the required minimum rate. As the minimum rate requirement becomes more stringent, more resources must be allocated to accommodate UTs experiencing inferior channel conditions. This results in a decline in the average achievable sum rate for all considered schemes, especially for those exploiting only S-CSIT. On the other hand, while the feasibility ratio for the S-CSIT-GMM and S-CSIT-UPA schemes declines sharply as the required minimum rate increases, the proposed GMM, the proposed PMM, and the Decoupled-GMM schemes manage to maintain a feasibility ratio close to  $100\%$. This observation highlights the benefits of the proposed two-timescale MA-enhanced MIMO system design.
	
	
	\section{Conclusion}
	In this paper, we investigated a multiuser MIMO downlink system exploiting S-CSIT and I-CSIR, where the BS and UTs are equipped with multiple MAs. The average achievable sum rates were maximized for two different receive movement modes, namely, GMM and PMM, via two-timescale optimization exploiting I-CSIR for receive APV design and S-CSIT for transmit APV and covariance matrix design. For GMM, the considered two-timescale problem was first decomposed into a series of short-term problems and one long-term problem. Then, a GA algorithm was proposed to obtain suboptimal receive APVs for the short-term problems for given I-CSIR samples. Based on the output of the GA algorithm, a series of convex objective/feasibility surrogates for the long-term problem were constructed and solved utilizing the CSSCA algorithm. Furthermore, we introduced the PMM for the receive MAs to facilitate efficient antenna movement and the development of the PDD-SSCA algorithm, which can acquire a KKT solution for the resulting two-timescale stochastic optimization problem almost surely. Numerical results revealed that, for GMM and PMM, the proposed MA-enhanced systems exploiting S-CSIT and I-CSIR can significantly improve the average achievable sum rates and the feasibility ratio compared to the conventional UPA system as well as an MA-enhanced system with GMM but exploiting only S-CSIT. Moreover, the proposed GMM scheme achieved the best achievable rate performance for all considered simulation settings. For future work, to minimize the movement of the receive MAs, the solution nearest to the previous position can be selected among multiple solutions that yield the same phase for each channel path.
	
	\begin{appendices}
		\section{  }
		\label{Ap: gradient rate}
		According to equation $\ln\left(\operatorname{det}\left(\mathbf{Z}\right)\right)=\operatorname{tr}\left\{\mathbf{Z}^{-1}\partial \mathbf{Z}\right\}$ in \cite{CVMD}, the partial derivative of $R_k\left(\mathbf{t}, \mathbf{r}_k,\left\{\mathbf{Q}_i\right\},\mathbf{H}_k\right)$ w.r.t. $\mathbf{Q}_i$ can be derived as shown in \eqref{eq: partial Rk Qi} at the top of the next page.
		\begin{figure*}[t]
			\begin{equation}
				\label{eq: partial Rk Qi}
				\begin{aligned}
					\frac{\partial}{\partial \mathbf{Q}_i} R_k
					=& \frac{\partial}{\partial \mathbf{Q}_i} \operatorname{tr} \left\{  \frac{1}{\ln 2} \left( \sigma^2 \mathbf{I}_{M}+\mathbf{H}_k \sum\nolimits_{i=1}^K\mathbf{Q}_i \mathbf{H}_k^{H} \right)^{-1} \mathbf{H}_k \sum\nolimits_{i=1}^K\partial \mathbf{Q}_i \mathbf{H}_k^{H} \right\} \\
					-& \frac{\partial}{\partial \mathbf{Q}_i} \operatorname{tr} \left\{  \frac{1}{\ln 2} \left( \sigma^2 \mathbf{I}_{M}+\mathbf{H}_k \sum\nolimits_{i\neq k}\mathbf{Q}_i \mathbf{H}_k^{H} \right)^{-1} \mathbf{H}_k \sum\nolimits_{i\neq k}\partial \mathbf{Q}_i \mathbf{H}_k^{H} \right\} \\
					=&
					\begin{cases}
						{\mathbf{A}_{Q_i,k}\triangleq\frac{1}{\ln 2} \mathbf{H}_k^{H}  \left( \sigma^2 \mathbf{I}_{M}+\mathbf{H}_k \sum_{i=1}^K\mathbf{Q}_i \mathbf{H}_k^{H} \right)^{-1} \mathbf{H}_k,} & \text { if } i=k, \\ 
						{\mathbf{A}_{Q_i,k}-\frac{1}{\ln 2} \mathbf{H}_k^{H} \left(  \sigma^2 \mathbf{I}_{M}+\mathbf{H}_k \sum_{i\neq k}\mathbf{Q}_i \mathbf{H}_k^{H} \right)^{-1} \mathbf{H}_k,} & \text { if } i\neq k.
					\end{cases}
				\end{aligned}
			\end{equation}
			\hrule
		\end{figure*}
		Similarly, the partial derivative of $R_k\left(\mathbf{t}, \mathbf{r}_k,\left\{\mathbf{Q}_i\right\},\mathbf{H}_k\right)$ w.r.t. $\mathbf{H}_k$ can be written as
		\begin{equation}
			\label{eq: partial Rk Hk}
			\begin{aligned}
				\frac{\partial}{\partial \mathbf{H}_k} R_k
				=& \frac{\partial}{\partial \mathbf{H}_k} \operatorname{tr} \left\{ \operatorname{Re} \left\{ \left( \mathbf{A}_{k,1} - \mathbf{A}_{k,2} \right) \partial \mathbf{H}_k \right\} \right\} \\
			\end{aligned}
		\end{equation}
		with matrices
		\begin{equation}
			\begin{aligned}
				\mathbf{A}_{k,1} &= \frac{2}{\ln 2} \sum_{i=1}^K\mathbf{Q}_i \mathbf{H}_k^{H} \left( \sigma^2 \mathbf{I}_{M}+\mathbf{H}_k \sum_{i=1}^K\mathbf{Q}_i \mathbf{H}_k^{H} \right)^{-1},\\
				\mathbf{A}_{k,2} &= \frac{2}{\ln 2} \sum_{i\neq k}\mathbf{Q}_i \mathbf{H}_k^{H} \left( \sigma^2 \mathbf{I}_{M}+\mathbf{H}_k \sum_{i\neq k}\mathbf{Q}_i \mathbf{H}_k^{H} \right)^{-1}.
			\end{aligned}
		\end{equation}
		Moreover, we define the differential vectors
		\begin{equation}
			\begin{aligned}
				\partial \mathbf{x}_{r,k} \triangleq & \left( \partial 	x_{r,k,1}, \partial x_{r,k,2}, \cdots, \partial x_{r,k,M} \right)^{T}, \\
				\partial \mathbf{y}_{r,k} \triangleq & \left( \partial 	y_{r,k,1}, \partial y_{r,k,2}, \cdots, \partial y_{r,k,M} \right)^{T}, \\
				\partial \mathbf{x}_{t} \triangleq & \left( \partial 	x_{t,1}, \partial x_{t,2}, \cdots, \partial x_{t,N} \right)^{T}, \\ 
				\partial \mathbf{y}_{t} \triangleq & \left( \partial 	y_{t,1}, \partial y_{t,2}, \cdots, \partial y_{t,N} \right)^{T},
			\end{aligned}
		\end{equation}
		and matrices 
		\begin{equation}
			\label{eq: Delta}
			\begin{aligned}
				\mathbf{\Delta}_{x_r,k} \triangleq \operatorname{diag}&\left\{ \frac{-j2\pi}{\lambda} \left( \sin\theta_{r,k}^{1}\cos\phi_{r,k}^{1},\right.\right.\\&\left.\left.\cdots,\sin\theta_{r,k}^{L_r}\cos\phi_{r,k}^{L_r} \right)^T\right\}, \\
				\mathbf{\Delta}_{y_r,k} \triangleq \operatorname{diag}&\left\{ \frac{-j2\pi}{\lambda} \left(
				\cos\theta_{r,k}^{1},\cdots,\cos\theta_{r,k}^{L_r} \right)^T\right\}, \\ 
				\mathbf{\Delta}_{x_t,k} \triangleq \operatorname{diag}&\left\{ \frac{j2\pi}{\lambda} \left( \sin\theta_{t,k}^{1}\cos\phi_{t,k}^{1},\right.\right.\\&\left.\left.\cdots,\sin\theta_{t,k}^{L_t}\cos\phi_{t,k}^{L_t} \right)^T\right\}, \\
				\mathbf{\Delta}_{y_t,k} \triangleq \operatorname{diag}&\left\{ \frac{j2\pi}{\lambda} \left( \cos\theta_{t,k}^{1},\cdots,\cos\theta_{t,k}^{L_t} \right)^T\right\}.
			\end{aligned}
		\end{equation}
		Then, the differential of $\mathbf{H}_k$ is given by
		\begin{equation}
			\label{eq: channel matrix differential}
			\begin{aligned}
				\partial \mathbf{H}_k 
				=& \operatorname{diag} \left\{ \partial \mathbf{x}_{r,k} \right\} \mathbf{B}_{x_r,k} + \operatorname{diag} \left\{ \partial \mathbf{y}_{r,k} \right\} \mathbf{B}_{y_r,k} \\
				+& \mathbf{B}_{x_t,k} \operatorname{diag} \left\{ \partial \mathbf{x}_{t} \right\} + \mathbf{B}_{y_t,k} \operatorname{diag} \left\{ \partial \mathbf{y}_{t} \right\},
			\end{aligned}
		\end{equation}
		where 
		\begin{equation}
			\begin{aligned}
				\mathbf{B}_{x_r,k} = & \mathbf{F}_{k}^H\left(\mathbf{r}_k\right) \mathbf{\Delta}_{x_r,k} \mathbf{\Sigma}_k \mathbf{G}_{k}\left(\mathbf{t}\right), \\
				\mathbf{B}_{y_r,k} = & \mathbf{F}_{k}^H\left(\mathbf{r}_k\right) \mathbf{\Delta}_{y_r,k} \mathbf{\Sigma}_k \mathbf{G}_{k}\left(\mathbf{t}\right), \\
				\mathbf{B}_{x_t,k} = & \mathbf{F}_{k}^H\left(\mathbf{r}_k\right) \mathbf{\Sigma}_k \mathbf{\Delta}_{x_t,k} \mathbf{G}_{k}\left(\mathbf{t}\right), \\
				\mathbf{B}_{y_t,k} = & \mathbf{F}_{k}^H\left(\mathbf{r}_k\right) \mathbf{\Sigma}_k \mathbf{\Delta}_{y_t,k} \mathbf{G}_{k}\left(\mathbf{t}\right).
			\end{aligned}
		\end{equation}
		Combining \eqref{eq: partial Rk Hk} and \eqref{eq: channel matrix differential}, the partial derivatives of $R_k\left(\mathbf{t}, 	\mathbf{r}_k,\left\{\mathbf{Q}_i\right\},\mathbf{H}_k\right)$ w.r.t. $\mathbf{x}_{t}$ and $\mathbf{y}_{t}$ are given by
		\begin{equation}
			\begin{aligned}
				\nabla_{\mathbf{x}_{t}} R_k
				=& \operatorname{diag}\left\{ \operatorname{Re} \left\{ \left( \mathbf{A}_{k,1} - \mathbf{A}_{k,2} \right) \mathbf{B}_{x_t,k} \right\} \right\}, \\
				\nabla_{\mathbf{y}_{t}} R_k
				=& \operatorname{diag}\left\{ \operatorname{Re} \left\{ \left( \mathbf{A}_{k,1} - \mathbf{A}_{k,2} \right) \mathbf{B}_{y_t,k} \right\} \right\},
			\end{aligned}
		\end{equation}
		which leads to $\nabla_{\mathbf{t}} R_k\left(\mathbf{t}, \mathbf{r}_k,\left\{\mathbf{Q}_i\right\},\mathbf{H}_k\right)$ by adjusting the  element positions.
		Similarly, the partial derivatives of $R_k\left(\mathbf{t}, 	\mathbf{r}_k,\left\{\mathbf{Q}_i\right\},\mathbf{H}_k\right)$ w.r.t. $\mathbf{r}_k$ can be constructed from
		\begin{equation}
			\begin{aligned}
				\label{eq: rate differential r}
				\nabla_{\mathbf{x}_{x_r,k}} R_k = &  \operatorname{diag}\left\{ \operatorname{Re} \left\{ \mathbf{B}_{x_r,k} \left( \mathbf{A}_{k,1} - \mathbf{A}_{k,2} \right) \right\} \right\}, \\
				\nabla_{\mathbf{x}_{y_r,k}} R_k = &  \operatorname{diag}\left\{ \operatorname{Re} \left\{ \mathbf{B}_{y_r,k} \left( \mathbf{A}_{k,1} - \mathbf{A}_{k,2} \right) \right\} \right\}.
			\end{aligned}
		\end{equation}
		
	\end{appendices}


\begin{thebibliography}{00}
		\bibliographystyle{unsrt}
		
		
		\bibitem{CLoM}A. Goldsmith, S. A. Jafar, N. Jindal, and S. Vishwanath, ``Capacity limits of MIMO channels,''  \emph{IEEE J. Sel. Areas Commun.}, vol. 21, no. 5, pp. 684-702, Jun. 2003.
		
		\bibitem{Nlaa}P. Wang, Y. Li, Y. Peng, S. C. Liew, and B. Vucetic, ``Non-uniform linear antenna array design and optimization for millimeter-wave communications,''  \emph{IEEE Trans. Wireless Commun.}, vol. 15, no. 11, pp. 7343-7356, Nov. 2016.
		
		\bibitem{Nadf2}M. Palaiologos, M. H. C. García, A. Kakkavas, R. A. Stirling-Gallacher, and G. Caire, ``Non-uniform array design for robust LoS MIMO via convex optimization,'' \emph{Annual International Symposium on Personal, Indoor and Mobile Radio Commun. (PIMRC)}, Sep. 2023.
		
		\bibitem{BLIF} K. K. Wong, K. F. Tong, Y. Shen, Y. Chen, and Y. Zhang, ``Bruce Lee-inspired fluid antenna system: Six research topics and the potentials for 6G,'' \emph{Frontiers Commun. Netw.}, 3:853416, Mar. 2022.
		
		
		\bibitem{FASN}W. K. New, K. K. Wong, X. Hao, K. F. Tong, and C. B. Chae, ``Fluid antenna system: New insights on outage probability and diversity gain,'' \emph{IEEE Trans. Wireless Commun.}, vol. 23, no. 1, pp. 128-140, Jan. 2024.
		
		\bibitem{FAwL} C. Skouroumounis and I. Krikidis, ``Fluid antenna with linear MMSE channel estimation for large-scale cellular networks,'' \emph{IEEE Trans. Commun.},  vol. 71, no. 2, pp. 1112-1125, Feb. 2022.
		
		\bibitem{Mapa} L. Zhu, W. Ma, and R. Zhang, ``Modeling and performance analysis for movable antenna enabled wireless communications,'' \emph{IEEE Trans. Wireless Commun.},  vol. 23, no. 6, pp. 6234-6250, Jun. 2024.
		
		\bibitem{Mccf} W. Ma, L. Zhu, and R. Zhang, ``MIMO capacity characterization for movable antenna systems,'' \emph{IEEE Trans. Wireless Commun.}, vol. 23, no. 4, pp. 3392-3407, Apr. 2024.
		
		\bibitem{JBaA} X. Chen, B. Feng, Y. Wu, D. W. K. Ng and R. Schober, ``Joint beamforming and antenna movement design for moveable antenna systems based on Statistical CSI,'' \emph{IEEE Global Commun. Conf.}, Dec. 2023.
		
		\bibitem{Eemf} X. Chen, B. Feng, Y. Wu, and W. Zhang, ``Energy efficiency maximization for movable antenna-enhanced system based on statistical CSI,''  \emph{IEEE Intern. Conf. Commun.}, Jun. 2025.
		
		\bibitem{WSRM} B. Feng, Y. Wu, X.-G. Xia , and C. Xiao, ``Weighted sum-rate maximization for movable antenna-enhanced wireless networks,'' \emph{IEEE Wireless Commun. Lett.}, vol. 13, no. 6, pp. 1770-1774, Jun. 2024.
		
		\bibitem{MAAE} L. Zhu, W. Ma, and R. Zhang, ``Movable-antenna array enhanced beamforming: Achieving full array gain with null steering,'' \emph{IEEE Commun. Lett.}, vol. 27, no. 12, pp. 3340-3344, Oct. 2023.
		
		\bibitem{MBFw} W. Ma, L. Zhu, and R. Zhang, ``Multi-beam forming with movable-antenna array,'' \emph{IEEE Commun. Lett.}, vol. 28, no. 3, pp. 697-701, Jan. 2024.
		
		\bibitem{MEMC} L. Zhu, W. Ma, B. Ning, and R. Zhang, ``Movable-antenna enhanced multiuser communication via antenna position optimization,'' \emph{IEEE Trans. Wireless Commun.}, vol. 23, no. 7, pp. 7214-7229, Jul. 2024.
		
		
		\bibitem{MAEM} Y. Wu, D. Xu, D. W. K. Ng, W. Gerstacker, and R. Schober, ``Movable antenna-enhanced multiuser communication: Optimal discrete antenna positioning and beamforming,'' \emph{IEEE Global Commun. Conf.}, Dec. 2023.
		
		\bibitem{APaB} H. Qin, W. Chen, Z. Li, Q. Wu, N. Cheng, and F. Chen, ``Antenna positioning and beamforming design for movable-antenna enabled multi-user downlink communications,'' arXiv preprint arXiv:2311.03046v1, 2023.
		
		\bibitem{SRMf} Z. Cheng, N. Li, J. Zhu, X. She, C. Ouyang, and P. Chen, ``Sum-rate maximization for movable antenna enabled multiuser communications,'' arXiv preprint arXiv:2309.11135v1, 2023.
		
		\bibitem{MCwMZXLX} Z. Xiao, X. Pi, L. Zhu, X. G. Xia, and R. Zhang, ``Multiuser communications with movable-antenna base station: Joint antenna positioning, receive combining, and power control,'' \emph{IEEE Trans. Wireless Commun.}, vol. 23, no. 12, pp. 19744-19759, Dec. 2024.
		
		\bibitem{CMfF} H. Xu, K. K. Wong, W. K. New, G. Zhou, R. Murch, C. B. Chae, Y. Zhu, and S. Jin, ``Capacity maximization for FAS-assisted multiple access channels,'' \emph{IEEE Trans. Commun.}, early access, Dec. 2024.
		
		\bibitem{MAEA} Z. Cheng, N. Li, J. Zhu, X. She, C. Ouyang, and P. Chen, ``Movable antenna-empowered AirComp,'' arXiv preprint arXiv:2309.12596v1, 2023.
		
		\bibitem{MAAEGQJK} G. Hu, Q. Wu, J. Ouyang, K. Xu, Y. Cai, and N. A. Dhahir, ``Movable-antenna array-enabled wireless communication with CoMP reception,'' \emph{IEEE Commun. Lett.}, vol. 28, no. 4, pp. 947-951, Apr. 2024.
		
		\bibitem{SWCv} G. Hu, Q. Wu, K. Xu, J. Si, and N. A. Dhahir, ``Secure wireless communication via movable-antenna array,''  \emph{IEEE Signal Process. Lett.}, vol. 31, pp. 516-520, Jan. 2024.
		
		\bibitem{CSBC} W. Ma, L. Zhu, and R. Zhang, ``Compressed sensing based channel estimation for  movable antenna communications,'' \emph{IEEE Commun. Lett.}, vol. 27, no. 10, pp. 2747-2751, Oct. 2023.
		
		\bibitem{CEfM} Z. Xiao, S. Cao, L. Zhu, Y. Liu, X. G. Xia, and R. Zhang, ``Channel estimation for movable antenna communication systems: A framework based on compressed sensing,'' \emph{IEEE Trans. Wireless Commun.}, vol. 23, no. 9, pp. 11814 - 11830, Sep. 2024.
		
		
		\bibitem{CSTC} M. A. Maddah-Ali and D. Tse, “Completely stale transmitter channel state information is still very useful,” \emph{IEEE Trans. Inf. Theory}, vol. 58, no. 7, pp. 4418–4431, Jul. 2012.
		
		\bibitem{MBCW} N. Jindal, ``MIMO broadcast channels with finite-rate feedback,'' \emph{IEEE Trans. Info. Theory}, vol. 52, no. 11, pp. 5045–5060, Nov. 2006.
		
		
		
		\bibitem{FmMv} M. Barzegar Khalilsarai, S. Haghighatshoar, X. Yi, and G. Caire, ``FDD massive MIMO via UL/DL channel covariance
		extrapolation and active channel sparsification,'' \emph{IEEE Trans. Wireless Commun.}, vol. 18, no. 1, pp. 121–135, Jan. 2019.
		
		\bibitem{TMCb} K. B. Petersen and M. S. Pedersen, \emph{The Matrix Cookbook}. Kgs. Lyngby, Denmark: Tech. Univ. Denmark, 2006.
		
		
		
		
		
		
		\bibitem{AAfL}D. Pinchera, M. D. Migliore, F. Schettino, and G. Panariello, ``Antenna arrays for line-of-sight massive MIMO: Half wavelength is not enough,''  \emph{Electronics}, vol. 6, no. 3, pp. 57-57, Sep. 2017.
		
		\bibitem{FAAM} Y. Ye, L. You, J. Wang, H. Xu, K. K. Wong, and X. Gao, ``Fluid antenna-assisted MIMO transmission exploiting statistical CSI,'' \emph{IEEE Commun. Lett.}, vol. 28, no. 1, pp. 223-227, Nov. 2023.
		
		
		
		\bibitem{SSCA} A. Liu, V. K. N. Lau, and B. Kananian, ``Stochastic successive convex approximation for nonconvex constrained stochastic optimization,''  \emph{IEEE Trans. Signal Process.}, vol. 67, no. 16, pp. 4189-4203, Aug. 2019.
		
		\bibitem{MDCW}J. R. Magnus and H. Neudecker, \emph{Matrix Differential Calculus With Applications in Statistics and Econometrics}. Hoboken, NJ, USA: Wiley, 1995.
		
		\bibitem{Aito}E. K. Chong and S. H. Zak, \emph{An Introduction to Optimization}. Hoboken, New Jersey, USA: John Wiley \& Sons, 2013.
		
		\bibitem{CO}S. Boyd and L. Vandenberghe, \emph{Convex Optimization}. Cambridge, U.K.: Cambridge Univ. Press, 2004.
		
		
		\bibitem{Np} R. Cottle, \emph{Nonlinear Programming}. Hoboken, New Jersey, USA: John Wiley \& Sons, 1972.
		
		
		
		\bibitem{FPMf} J. Zheng, J. Zhang, H. Du, D. Niyato, S. Sun, Bo Ai, and K. B. Letaief, ``Flexible-position MIMO for wireless communications: fundamentals, challenges, and future directions,'' \emph{IEEE Wireless Commun.}, vol. 31, no. 5, pp. 18-26, Oct. 2023.
		
		
		
		
		
		
		\bibitem{MMAi} Y. Sun, P. Babu, and D. P. Palomar, ``Majorization-minimization algorithms in signal processing, communications, and machine learning,'' \emph{IEEE Trans. Signal Process.}, vol. 65, no. 3, pp. 794-816, Feb. 2017.
		
		\bibitem{TSSO} A. Liu, R. Yang, T. Q. S. Quek, and M. Zhao,``Two-stage stochastic optimization via primal-dual decomposition and deep unrolling,'' \emph{IEEE Trans. Signal Process.}, vol. 69, pp. 3000-3015, 2021.
		
		
		\bibitem{CVMD} A. Hjørungnes, and D. Gesbert, ``Complex-valued matrix differentiation: Techniques and key results,'' \emph{IEEE Trans. Signal Process.}, vol. 55, no. 6, pp. 2740 - 2746, Jun. 2007.
		
	\end{thebibliography}
\end{document}